# Spin and orbital angular momenta of electromagnetic waves in classical and quantum electrodynamics

Masud Mansuripur

James C. Wyant College of Optical Sciences, The University of Arizona, Tucson

**Abstract**. A plane, monochromatic electromagnetic wave propagating in free space can have a certain amount of spin angular momentum but cannot possess any orbital angular momentum. Even the spin angular momentum of the plane-wave is difficult to evaluate without resort to certain mathematical limit arguments. Both spin and orbital angular momenta can be computed for a wavepacket of finite duration and finite cross-sectional area using standard methods of classical electrodynamics. Extending these results to finite wavepackets in quantum electrodynamics requires subtle arguments in conjunction with the multimodal structure of the wavepacket. This paper presents some of the nuances of classical as well as quantum-optical methods for analyzing the spin and orbital angular momenta of electromagnetic waves.

**1. Introduction**. Quantization of the electromagnetic (EM) field starts with a reformulation of the classical Maxwell-Lorentz electrodynamics that enables a smooth transition to operator-based methods of quantum mechanics.[1] One typically assumes that a number of electric point-charges $q_n$, located in free space at their instantaneous positions $\boldsymbol{r}_n(t)$ and moving with velocities $\dot{\boldsymbol{r}}_n(t)$, produce an electric field $\boldsymbol{E}(\boldsymbol{r},t)$ and a magnetic field $\boldsymbol{B}(\boldsymbol{r},t)$ at all points $(\boldsymbol{r},t)$ in space and time. The relations among the electric charge-density $\rho_{\text{free}}(\boldsymbol{r},t)=\sum_n q_n\delta(\boldsymbol{r}-\boldsymbol{r}_n)$, the electric current-density $\boldsymbol{J}_{\text{free}}(\boldsymbol{r},t)=\sum_n q_n\dot{\boldsymbol{r}}_n(t)\delta(\boldsymbol{r}-\boldsymbol{r}_n)$,[†] and the resulting $\boldsymbol{E}$ and $\boldsymbol{B}$ fields are dictated by Maxwell's classical electrodynamic equations, whose solutions are often facilitated by the introduction of a pair of auxiliary fields, namely, the scalar potential $\psi(\boldsymbol{r},t)$ and the vector potential $\boldsymbol{A}(\boldsymbol{r},t)$.[2,3] In general, $\boldsymbol{B}=\boldsymbol{\nabla}\times\boldsymbol{A}$ and $\boldsymbol{E}=-\boldsymbol{\nabla}\psi-\partial\boldsymbol{A}/\partial t$.[‡]

A well-known approach to solving partial differential equations, such as those of Maxwell, involves Fourier transformation.[1,3,4] In three-dimensional space, a function of $\boldsymbol{r}=(x,y,z)$ is transformed into a function of $\boldsymbol{k}=(k_x,k_y,k_z)$ via the Fourier kernel $\exp(\mathrm{i}\boldsymbol{k}\cdot\boldsymbol{r})$, which converts differentiation with respect to the spatial coordinates into multiplication by the corresponding $k$-vector components. For instance, if the Fourier transform of $\boldsymbol{E}(\boldsymbol{r},t)$ is denoted $\widetilde{\boldsymbol{E}}(\boldsymbol{k},t)$, then $\boldsymbol{\nabla}\cdot\boldsymbol{E}$ transforms into $\mathrm{i}\boldsymbol{k}\cdot\widetilde{\boldsymbol{E}}$. Similarly, $\boldsymbol{\nabla}\psi(\boldsymbol{r},t)$ becomes $\mathrm{i}\boldsymbol{k}\tilde{\psi}(\boldsymbol{k},t)$, while $\boldsymbol{\nabla}\times\boldsymbol{A}(\boldsymbol{r},t)$ transforms into $\mathrm{i}\boldsymbol{k}\times\widetilde{\boldsymbol{A}}(\boldsymbol{k},t)$. Thus, upon transformation to the $(\boldsymbol{k},t)$ space, Maxwell's equations become free of derivatives with respect to the $x,y,z$ spatial coordinates; the only remaining differentiation is then with respect to the time coordinate $t$, and the dynamic equations of Maxwell reduce to coupled first-order ordinary differential equations, resembling those of driven harmonic oscillators.[1]

Further simplifications arise from the splitting of $\widetilde{\boldsymbol{E}}(\boldsymbol{k},t)$ into a longitudinal component $\widetilde{\boldsymbol{E}}_\parallel(\boldsymbol{k},t)=[\hat{\boldsymbol{\kappa}}\cdot\widetilde{\boldsymbol{E}}(\boldsymbol{k},t)]\hat{\boldsymbol{\kappa}}$ and a transverse component $\widetilde{\boldsymbol{E}}_\perp(\boldsymbol{k},t)=\widetilde{\boldsymbol{E}}(\boldsymbol{k},t)-\widetilde{\boldsymbol{E}}_\parallel(\boldsymbol{k},t)$, where $\hat{\boldsymbol{\kappa}}=\boldsymbol{k}/k$ is the unit-vector along the direction of $\boldsymbol{k}$ in the Fourier domain.[1] As it turns out, the inverse Fourier transform of $\widetilde{\boldsymbol{E}}_{\parallel}(\boldsymbol{k},t)$, denoted $\boldsymbol{E}_{\parallel}(\boldsymbol{r},t)$, has a simple interpretation in the $(\boldsymbol{r},t)$ space, where it represents the Coulomb $E$-field produced by all the point-charges $q_n$ at their present locations $\boldsymbol{r}_n(t)$, without regard for their motions or for any retardation effects. More importantly, certain fractions of the overall EM field's energy, linear momentum, and angular momentum are tied to the longitudinal component $\boldsymbol{E}_{\parallel}(\boldsymbol{r},t)$ of the $E$-field; these can be split off and bundled together with the kinetic energies, mechanical linear momenta, and mechanical angular momenta of the point-charges that are the original sources of the EM fields.[1] In this way, the dynamics of the transverse EM fields, $\boldsymbol{E}_{\perp}(\boldsymbol{r},t)$ and $\boldsymbol{B}(\boldsymbol{r},t)$, become partially decoupled from the dynamics of the point-charges, namely, those of $\boldsymbol{r}_n(t)$ under the influence of the local Lorentz force $q_n\{\boldsymbol{E}[\boldsymbol{r}_n(t),t]+\dot{\boldsymbol{r}}_n(t)\times\boldsymbol{B}[\boldsymbol{r}_n(t),t]\}$. The sole driver of the transverse radiation field will then be $\boldsymbol{J}_{\perp(\text{free})}(\boldsymbol{r},t)$.

---

[†] The three-dimensional Dirac delta-function $\delta(\boldsymbol{r})=\delta(x)\delta(y)\delta(z)$ is used here to represent the charge and current densities of a localized point-charge. Each of the one-dimensional $\delta$-functions has the units of [1/meter] in the *SI* system of units, resulting in the units of $\rho_{\text{free}}(\boldsymbol{r},t)$ being [coulomb/m$^3$]. By the same token, the units of the current-density $\boldsymbol{J}_{\text{free}}(\boldsymbol{r},t)$ are [ampère/m$^2$], since the particle velocities $\dot{\boldsymbol{r}}_n(t)$ have the dimensions of [m/sec].

[‡] The definition of the vector potential $\boldsymbol{A}$ as the function whose curl equals the magnetic $B$-field guarantees the satisfaction of Maxwell's 4$^{\text{th}}$ equation, $\boldsymbol{\nabla}\cdot\boldsymbol{B}(\boldsymbol{r},t)=0$. Similarly, the definition of the scalar potential $\psi$ as the function that yields $\boldsymbol{E}=-\boldsymbol{\nabla}\psi-\partial\boldsymbol{A}/\partial t$ guarantees the satisfaction of Maxwell's 3$^{\text{rd}}$ equation, $\boldsymbol{\nabla}\times\boldsymbol{E}=-\partial\boldsymbol{B}/\partial t$.[1-5]

We begin in Sec.2 by showing that the longitudinal component $\boldsymbol{E}_{\parallel}(\boldsymbol{r},t)$ of the electric field can be expressed as the Coulomb $E$-field of all the point-charges $q_n$ without regard for retardation, as though, at time $t$, each point-charge had always been at its current location, $\boldsymbol{r}_n(t)$. The next three sections examine the energy, the linear momentum, and the angular momentum of the longitudinal $E$-field. In Sec.3, we show that the integrated energy content of $\boldsymbol{E}_{\parallel}(\boldsymbol{r},t)$ over the entire space consists of two contributions: (i) the sum of the self-energies of all the point-charges, each being infinite in magnitude, and (ii) the sum over the potential energies of all the point-charges, each experiencing the local and instantaneous scalar potential created by all the other charges. Similarly, the linear momentum of the EM field associated with $\boldsymbol{E}_{\parallel}(\boldsymbol{r},t)$ is shown in Sec.4 to be the sum over all the particles of the product of $q_n$ and the vector potential $\boldsymbol{A}(\boldsymbol{r},t)$ at the location $\boldsymbol{r}_n(t)$ of the particle, provided that the vector potential is evaluated in the Coulomb gauge. (An equivalent way of saying the same thing is that the transverse component $\boldsymbol{A}_{\perp}$ of the local vector potential must be used in calculating the EM linear momentum associated with the longitudinal $E$-field.) Finally, in Sec.5, we show that the angular momentum of the EM field associated with $\boldsymbol{E}_{\parallel}(\boldsymbol{r},t)$ is the sum over all the particles of $q_n\boldsymbol{r}_n(t)\times\boldsymbol{A}_{\perp}(\boldsymbol{r},t)$, where the vector potential is once again evaluated at the location $\boldsymbol{r}_n(t)$ of the corresponding particle. Thus, the two takeaways from the first part of the paper are that (i) in connection with the longitudinal $E$-field, the self-energies of the point-charges are infinite but constant and can, therefore, be ignored, and (ii) the EM energy, linear momentum, and angular momentum associated with $\boldsymbol{E}_{\parallel}(\boldsymbol{r},t)$ can be assigned to individual point-particles, since they depend exclusively on the values of $q_n$ and those of the scalar and vector potentials $\psi(\boldsymbol{r},t)$ and $\boldsymbol{A}(\boldsymbol{r},t)$ (albeit in the Coulomb gauge) at the instantaneous location $\boldsymbol{r}_n(t)$ of each particle.[1] Considering that the contributions of the longitudinal $E$-field to the system's energy and momenta can be bundled together with those of the material particles, there will be no need to include $\boldsymbol{E}_{\parallel}(\boldsymbol{r},t)$ in our subsequent discussion of the EM field and its quantization. Thus, the focus of the remaining part of the paper will be on the transverse field components $\boldsymbol{E}_{\perp}(\boldsymbol{r},t)$ and $\boldsymbol{B}(\boldsymbol{r},t)$, in conjunction with the transverse component $\boldsymbol{A}_{\perp}(\boldsymbol{r},t)$ of the vector potential.

Following a brief review of Maxwell's equations in the presence of point-charges in Sec.6, where we derive the relation between $\boldsymbol{A}_{\perp}(\boldsymbol{r},t)$ and the transverse component $\boldsymbol{J}_{\perp(\text{free})}(\boldsymbol{r},t)$ of the free current density, Sec.7 provides a detailed description of the so-called normal variables, which act as a point of departure from the classical theory into the quantum theory of electrodynamics.[1,4] The discretized propagating plane-wave modes of the EM field are introduced in this section, with each mode being specified by its frequency $\omega$, $k$-vector $\boldsymbol{k}$, the (generally complex-valued) unit-vector $\hat{\boldsymbol{e}}$ representing the mode's polarization, and the corresponding normal variable $\alpha(t)$ as well as its complex conjugate $\alpha^*(t)$. The normal variables $\alpha(t)$ and $\alpha^*(t)$ are the precursors of the annihilation and creation operators $\hat{a}$ and $\hat{a}^\dagger$ that play significant roles in the quantum theory of radiation, where they act on individual modes of the EM field to extract information from the occupying number states $|n\rangle$.[1,4,5]

In Sec.8 we relate the energy content as well as the linear momentum of each propagating $(\omega, \boldsymbol{k}, \hat{\boldsymbol{e}})$ mode in free space to the product of its $\alpha(t)$ and $\alpha^*(t)$ values. A similar treatment in Sec.9 leads to compact expressions for the spin and orbital angular momenta of propagating EM waves in free space. The final section, Sec.10, contains a brief summary as well as a few closing remarks. In the Appendix, we establish the equivalence of two different formulations of the optical angular momentum for free fields.

**2. Longitudinal $E$-field produced by a collection of point-charges in free space.** Consider a number of point-particles of charge $q_n$ moving along their individual trajectories $\boldsymbol{r}_n(t)$ in three-dimensional (3D) Euclidean space. The longitudinal $E$-field produced by these charges can be obtained directly from Maxwell's first equation, namely,[2,3]

$$\boldsymbol{\nabla}\cdot\varepsilon_0\boldsymbol{E}(\boldsymbol{r},t)=\rho_{\text{free}}(\boldsymbol{r},t)=\textstyle\sum_n q_n\delta[\boldsymbol{r}-\boldsymbol{r}_n(t)]. \quad \leftarrow \boxed{\varepsilon_0 \text{ is the permittivity of free space}} \tag{1}$$

Fourier transforming the above equation in 3D space, we find

$$\mathrm{i}\boldsymbol{k}\cdot\varepsilon_0\tilde{\boldsymbol{E}}(\boldsymbol{k},t)=\textstyle\sum_n q_n\iiint_{-\infty}^{\infty}\delta[\boldsymbol{r}-\boldsymbol{r}_n(t)]e^{-\mathrm{i}\boldsymbol{k}\cdot\boldsymbol{r}}\mathrm{d}\boldsymbol{r}=\sum_n q_n e^{-\mathrm{i}\boldsymbol{k}\cdot\boldsymbol{r}_n(t)}. \tag{2}$$

In the Fourier domain, the longitudinal component $\tilde{\boldsymbol{E}}_\parallel(\boldsymbol{k},t)$ of the $E$-field is the projection of $\tilde{\boldsymbol{E}}(\boldsymbol{k},t)$ onto the local $k$-vector, i.e., $[\hat{\boldsymbol{\kappa}}\cdot\tilde{\boldsymbol{E}}(\boldsymbol{k},t)]\hat{\boldsymbol{\kappa}}$, where $\hat{\boldsymbol{\kappa}}=\boldsymbol{k}/k$. Inverse Fourier transformation now yields

$$\begin{aligned}
\boldsymbol{E}_\parallel(\boldsymbol{r},t) &= -\frac{\mathrm{i}}{(2\pi)^3\varepsilon_0}\textstyle\sum_n q_n\iiint_{-\infty}^{\infty}(\boldsymbol{k}/k^2)e^{-\mathrm{i}\boldsymbol{k}\cdot\boldsymbol{r}_n(t)}e^{\mathrm{i}\boldsymbol{k}\cdot\boldsymbol{r}}\mathrm{d}\boldsymbol{k} \quad \leftarrow \boxed{\mathrm{d}\boldsymbol{k} \text{ stands for } \mathrm{d}k_x\mathrm{d}k_y\mathrm{d}k_z}\\
&= -\frac{1}{(2\pi)^3\varepsilon_0}\textstyle\sum_n q_n\boldsymbol{\nabla}_r\iiint_{-\infty}^{\infty}k^{-2}e^{\mathrm{i}\boldsymbol{k}\cdot[\boldsymbol{r}-\boldsymbol{r}_n(t)]}\mathrm{d}\boldsymbol{k}\\
&= -\frac{1}{(2\pi)^3\varepsilon_0}\textstyle\sum_n q_n\boldsymbol{\nabla}_r\int_{k=0}^{\infty}\int_{\theta=0}^{\pi}2\pi\sin\theta\, e^{\mathrm{i}k|\boldsymbol{r}-\boldsymbol{r}_n(t)|\cos\theta}\mathrm{d}\theta\mathrm{d}k\\
&= -\frac{1}{(2\pi)^2\varepsilon_0}\textstyle\sum_n q_n\boldsymbol{\nabla}_r\displaystyle\int_{k=0}^{\infty}\frac{2\mathrm{i}\sin[k|\boldsymbol{r}-\boldsymbol{r}_n(t)|]}{\mathrm{i}k|\boldsymbol{r}-\boldsymbol{r}_n(t)|}\mathrm{d}k\\
&= -\frac{1}{2\pi^2\varepsilon_0}\textstyle\sum_n q_n\boldsymbol{\nabla}_r|\boldsymbol{r}-\boldsymbol{r}_n(t)|^{-1}\int_{x=0}^{\infty}(\sin x/x)\mathrm{d}x\\
&= -\frac{1}{4\pi\varepsilon_0}\textstyle\sum_n q_n\boldsymbol{\nabla}_r|\boldsymbol{r}-\boldsymbol{r}_n(t)|^{-1}=\frac{1}{4\pi\varepsilon_0}\sum_n\frac{q_n[\boldsymbol{r}-\boldsymbol{r}_n(t)]}{|\boldsymbol{r}-\boldsymbol{r}_n(t)|^3}.
\end{aligned} \tag{3}$$

---

**Digression**. Alternatively, the integral on the first line of Eq.(3) may be evaluated as follows:

$$\boxed{\int_{\theta=0}^{\pi}\sin\theta\cos\theta\, e^{\mathrm{i}\alpha\cos\theta}\mathrm{d}\theta = 2\mathrm{i}(\sin\alpha-\alpha\cos\alpha)/\alpha^2}$$

$$\begin{aligned}
\boldsymbol{E}_\parallel(\boldsymbol{r},t) &= \frac{1}{\mathrm{i}(2\pi)^3\varepsilon_0}\textstyle\sum_n q_n\iiint_{-\infty}^{\infty}(\boldsymbol{k}/k^2)e^{\mathrm{i}\boldsymbol{k}\cdot[\boldsymbol{r}-\boldsymbol{r}_n(t)]}\mathrm{d}\boldsymbol{k}\\
&= \frac{1}{\mathrm{i}(2\pi)^3\varepsilon_0}\sum_n\frac{q_n[\boldsymbol{r}-\boldsymbol{r}_n(t)]}{|\boldsymbol{r}-\boldsymbol{r}_n(t)|}\int_{k=0}^{\infty}\int_{\theta=0}^{\pi}2\pi k\sin\theta\cos\theta\, e^{\mathrm{i}k|\boldsymbol{r}-\boldsymbol{r}_n(t)|\cos\theta}\mathrm{d}\theta\mathrm{d}k\\
&= \frac{1}{2\pi^2\varepsilon_0}\sum_n\frac{q_n[\boldsymbol{r}-\boldsymbol{r}_n(t)]}{|\boldsymbol{r}-\boldsymbol{r}_n(t)|}\int_{k=0}^{\infty}\frac{\sin[k|\boldsymbol{r}-\boldsymbol{r}_n(t)|]-k|\boldsymbol{r}-\boldsymbol{r}_n(t)|\cos[k|\boldsymbol{r}-\boldsymbol{r}_n(t)|]}{k|\boldsymbol{r}-\boldsymbol{r}_n(t)|^2}\mathrm{d}k\\
\boxed{\textstyle\int_0^\infty\cos x\,\mathrm{d}x \text{ is assumed to be zero}} \rightarrow &= \frac{1}{2\pi^2\varepsilon_0}\sum_n\frac{q_n[\boldsymbol{r}-\boldsymbol{r}_n(t)]}{|\boldsymbol{r}-\boldsymbol{r}_n(t)|^3}\int_{x=0}^{\infty}\frac{\sin(x)-x\cos(x)}{x}\mathrm{d}x=\frac{1}{4\pi\varepsilon_0}\sum_n\frac{q_n[\boldsymbol{r}-\boldsymbol{r}_n(t)]}{|\boldsymbol{r}-\boldsymbol{r}_n(t)|^3}.
\end{aligned} \tag{4}$$

---

Note that the longitudinal $E$-field turns out to be the same as the instantaneous $E$-field at the observation point $\boldsymbol{r}$ produced by all the charges if the charges happen to be static. In other words, the contributions of the charges to the $E$-field do not take retardation into account. However, the transverse component $\boldsymbol{E}_\perp$ of the $E$-field similarly fails to properly account for the effects of retardation in such a way that the total field $\boldsymbol{E}_\parallel + \boldsymbol{E}_\perp$ ends up being causal.

The longitudinal $E$-field is seen to be $\boldsymbol{E}_\parallel(\boldsymbol{r},t) = -\boldsymbol{\nabla}\psi(\boldsymbol{r},t) = -\boldsymbol{\nabla}\sum_n q_n/[4\pi\varepsilon_0|\boldsymbol{r}-\boldsymbol{r}_n(t)|]$, where $\psi(\boldsymbol{r},t)$ is the scalar potential in the Coulomb gauge. As always, the transverse $E$-field will be given by $\boldsymbol{E}_\perp(\boldsymbol{r},t) = -\partial\boldsymbol{A}_\perp(\boldsymbol{r},t)/\partial t$, where $\boldsymbol{A}_\perp(\boldsymbol{r},t) = \boldsymbol{A}(\boldsymbol{r},t)$ in the Coulomb gauge.

**3. Electromagnetic energy contained in the longitudinal component of the *E*-field**. The $E$-field energy-density in free space is generally given by $\tfrac{1}{2}\varepsilon_0|\boldsymbol{E}(\boldsymbol{r},t)|^2$. Integrating over the entire 3D space the energy-density of $\boldsymbol{E}_\parallel(\boldsymbol{r},t)$ as given by the first line in Eq.(3), we find

$$\mathcal{E}_\parallel(t) = \tfrac{1}{2}\varepsilon_0\iiint_{-\infty}^{\infty}\boldsymbol{E}_\parallel(\boldsymbol{r},t)\cdot\boldsymbol{E}_\parallel(\boldsymbol{r},t)\mathrm{d}\boldsymbol{r} \quad \leftarrow \boxed{\mathrm{d}\boldsymbol{r}\text{ stands for }\mathrm{d}x\mathrm{d}y\mathrm{d}z} \quad \boxed{\text{integration over }\boldsymbol{r}\text{ yields }(2\pi)^3\delta(\boldsymbol{k}+\boldsymbol{k}')}$$

$$= -\frac{1}{2(2\pi)^6\varepsilon_0}\sum_n\sum_m q_n q_m \iiint_{-\infty}^{\infty}(\boldsymbol{k}\cdot\boldsymbol{k}'/k^2k'^2)e^{-\mathrm{i}\boldsymbol{k}\cdot\boldsymbol{r}_n(t)}e^{-\mathrm{i}\boldsymbol{k}'\cdot\boldsymbol{r}_m(t)}e^{\mathrm{i}(\boldsymbol{k}+\boldsymbol{k}')\cdot\boldsymbol{r}}\mathrm{d}\boldsymbol{k}\mathrm{d}\boldsymbol{k}'\mathrm{d}\boldsymbol{r}$$

$$= -\frac{1}{2(2\pi)^3\varepsilon_0}\sum_n\sum_m q_n q_m \iiint_{-\infty}^{\infty}(\boldsymbol{k}\cdot\boldsymbol{k}'/k^2k'^2)e^{-\mathrm{i}\boldsymbol{k}\cdot\boldsymbol{r}_n(t)}e^{-\mathrm{i}\boldsymbol{k}'\cdot\boldsymbol{r}_m(t)}\delta(\boldsymbol{k}+\boldsymbol{k}')\mathrm{d}\boldsymbol{k}\mathrm{d}\boldsymbol{k}'$$

$$= \frac{1}{2(2\pi)^3\varepsilon_0}\sum_n\sum_m q_n q_m \iiint_{-\infty}^{\infty}(1/k^2)e^{-\mathrm{i}\boldsymbol{k}\cdot\boldsymbol{r}_n(t)}e^{\mathrm{i}\boldsymbol{k}\cdot\boldsymbol{r}_m(t)}\mathrm{d}\boldsymbol{k} \quad \leftarrow \boxed{\begin{array}{c}\text{sifting property of }\delta(\boldsymbol{k}+\boldsymbol{k}')\\ \text{replaces }\boldsymbol{k}'\text{ everywhere by }-\boldsymbol{k}\end{array}}$$

$$= \frac{1}{(2\pi)^2\varepsilon_0}\sum_n q_n^2\int_{k=0}^{\infty}\mathrm{d}k + \frac{1}{2(2\pi)^2\varepsilon_0}\sum_n\sum_{m\neq n} q_n q_m\int_{k=0}^{\infty}\int_{\theta=0}^{\pi}\sin\theta\, e^{\mathrm{i}k|\boldsymbol{r}_m(t)-\boldsymbol{r}_n(t)|\cos\theta}\mathrm{d}\theta\mathrm{d}k$$

$$= \frac{1}{(2\pi)^2\varepsilon_0}\sum_n q_n^2\int_{k=0}^{\infty}\mathrm{d}k + \frac{1}{(2\pi)^2\varepsilon_0}\sum_n\sum_{m\neq n} q_n q_m\int_{k=0}^{\infty}\frac{\sin[k|\boldsymbol{r}_m(t)-\boldsymbol{r}_n(t)|]}{k|\boldsymbol{r}_m(t)-\boldsymbol{r}_n(t)|}\mathrm{d}k$$

$$= \frac{1}{4\pi^2\varepsilon_0}\sum_n q_n^2\int_{k=0}^{\infty}\mathrm{d}k + \frac{1}{8\pi\varepsilon_0}\sum_n\sum_{m\neq n} q_n q_m/|\boldsymbol{r}_m(t)-\boldsymbol{r}_n(t)|. \tag{5}$$

In the above equation, the first term (representing the sum of the self-energies of all the point charges) is infinite, while the second term represents the sum of the potential energies of each point-charge within the (instantaneous) scalar potential produced by all the other charges.

**4. Contribution of longitudinal *E*-field to electromagnetic momentum**. The EM momentum at any given instant of time $t$ is the integral of $\varepsilon_0\boldsymbol{E}(\boldsymbol{r},t)\times\boldsymbol{B}(\boldsymbol{r},t)$ over the entire 3D space. Invoking the identity $\boldsymbol{B}(\boldsymbol{r},t) = \boldsymbol{\nabla}\times\boldsymbol{A}(\boldsymbol{r},t)$, the contribution of the longitudinal component $\boldsymbol{E}_\parallel$ of the $E$-field to the overall linear momentum of the EM field is evaluated as follows:

$$\boldsymbol{p}_\parallel^{(\mathrm{EM})}(t) = \varepsilon_0\iiint_{-\infty}^{\infty}\boldsymbol{E}_\parallel(\boldsymbol{r},t)\times\boldsymbol{B}(\boldsymbol{r},t)\mathrm{d}\boldsymbol{r} = \varepsilon_0\iiint_{-\infty}^{\infty}\boldsymbol{E}_\parallel(\boldsymbol{r},t)\times[\boldsymbol{\nabla}\times\boldsymbol{A}(\boldsymbol{r},t)]\mathrm{d}\boldsymbol{r}$$

$$= \varepsilon_0\iiint_{-\infty}^{\infty}[\boldsymbol{\nabla}(\boldsymbol{A}\cdot\boldsymbol{E}_\parallel) - (\boldsymbol{A}\cdot\boldsymbol{\nabla})\boldsymbol{E}_\parallel - (\boldsymbol{E}_\parallel\cdot\boldsymbol{\nabla})\boldsymbol{A} - \boldsymbol{A}\times(\underbrace{\boldsymbol{\nabla}\times\boldsymbol{E}_\parallel}_{0})]\mathrm{d}\boldsymbol{r} \quad \leftarrow \boxed{\text{standard vector identity}}$$

$$= \varepsilon_0\{\hat{\boldsymbol{x}}\iint_{-\infty}^{\infty}\underbrace{(\boldsymbol{A}\cdot\boldsymbol{E}_\parallel)|_{x=-\infty}^{\infty}}_{0}\mathrm{d}y\mathrm{d}z + \hat{\boldsymbol{y}}\iint_{-\infty}^{\infty}\underbrace{(\boldsymbol{A}\cdot\boldsymbol{E}_\parallel)|_{y=-\infty}^{\infty}}_{0}\mathrm{d}x\mathrm{d}z + \hat{\boldsymbol{z}}\iint_{-\infty}^{\infty}\underbrace{(\boldsymbol{A}\cdot\boldsymbol{E}_\parallel)|_{z=-\infty}^{\infty}}_{0}\mathrm{d}x\mathrm{d}y\}$$

$$-\varepsilon_0\{\iint_{-\infty}^{\infty}\underbrace{(A_x\boldsymbol{E}_\parallel)|_{x=-\infty}^{\infty}}_{0}\mathrm{d}y\mathrm{d}z + \iint_{-\infty}^{\infty}\underbrace{(A_y\boldsymbol{E}_\parallel)|_{y=-\infty}^{\infty}}_{0}\mathrm{d}x\mathrm{d}z + \iint_{-\infty}^{\infty}\underbrace{(A_z\boldsymbol{E}_\parallel)|_{z=-\infty}^{\infty}}_{0}\mathrm{d}x\mathrm{d}y$$

$$-\iiint_{-\infty}^{\infty}(\partial_x A_x + \partial_y A_y + \partial_z A_z)\boldsymbol{E}_\parallel\mathrm{d}\boldsymbol{r}\} \quad \leftarrow \boxed{\text{integration by parts}}$$

$$-\varepsilon_0\Big\{\iint_{-\infty}^{\infty}\cancelto{0}{(E_{\parallel x}\boldsymbol{A})}\Big|_{x=-\infty}^{\infty}\mathrm{d}y\mathrm{d}z+\iint_{-\infty}^{\infty}\cancelto{0}{(E_{\parallel y}\boldsymbol{A})}\Big|_{y=-\infty}^{\infty}\mathrm{d}x\mathrm{d}z+\iint_{-\infty}^{\infty}\cancelto{0}{(E_{\parallel z}\boldsymbol{A})}\Big|_{z=-\infty}^{\infty}\mathrm{d}x\mathrm{d}y$$

$$-\iiint_{-\infty}^{\infty}(\partial_x E_{\parallel x}+\partial_y E_{\parallel y}+\partial_z E_{\parallel z})\boldsymbol{A}\mathrm{d}\boldsymbol{r}\Big\}\quad\leftarrow\boxed{\text{integration by parts}}$$

$$=\varepsilon_0\iiint_{-\infty}^{\infty}\cancelto{0}{(\boldsymbol{\nabla}\cdot\boldsymbol{A})}\boldsymbol{E}_{\parallel}\mathrm{d}\boldsymbol{r}+\iiint_{-\infty}^{\infty}(\varepsilon_0\boldsymbol{\nabla}\cdot\boldsymbol{E}_{\parallel})\boldsymbol{A}\mathrm{d}\boldsymbol{r}\quad\leftarrow\boxed{\text{Coulomb gauge}}$$

$$=\iiint_{-\infty}^{\infty}\{\textstyle\sum_n q_n\delta[\boldsymbol{r}-\boldsymbol{r}_n(t)]\}\boldsymbol{A}(\boldsymbol{r},t)\mathrm{d}\boldsymbol{r}=\sum_n q_n\boldsymbol{A}[\boldsymbol{r}_n(t),t].\quad\leftarrow\boxed{\text{Maxwell's 1}^{\text{st}}\text{ equation}}\qquad(6)$$

It is seen that the sum of the products of $q_n$ and the vector potential $\boldsymbol{A}(\boldsymbol{r},t)$ at the location $\boldsymbol{r}_n(t)$ of the point-charge $q_n$ yields the contribution of $\boldsymbol{E}_{\parallel}$ to the overall EM momentum, provided that $\boldsymbol{A}(\boldsymbol{r},t)$ is evaluated in the Coulomb gauge. As a matter of fact, restriction to Coulomb gauge would not be needed if we wrote $\boldsymbol{p}_{\parallel}^{(\text{EM})}(t)=\sum_n q_n\boldsymbol{A}_{\perp}[\boldsymbol{r}_n(t),t]$, since $\boldsymbol{A}_{\perp}$ is gauge-independent. This would also be obvious if we started the derivation of Eq.(6) by writing $\boldsymbol{B}=\boldsymbol{\nabla}\times\boldsymbol{A}_{\perp}$. Another way to see this is by noting that, in the penultimate line of Eq.(6), both integrals can be evaluated in the Fourier domain, where $\boldsymbol{\nabla}\cdot\boldsymbol{A}\rightarrow\mathrm{i}\boldsymbol{k}\cdot\widetilde{\boldsymbol{A}}=\mathrm{i}k\tilde{A}_{\parallel}$, $\varepsilon_0\boldsymbol{E}_{\parallel}\rightarrow-\mathrm{i}(\boldsymbol{k}/k^2)\tilde{\rho}_{\text{free}}$, $\varepsilon_0\boldsymbol{\nabla}\cdot\boldsymbol{E}_{\parallel}\rightarrow\tilde{\rho}_{\text{free}}$, and $\boldsymbol{A}\rightarrow\widetilde{\boldsymbol{A}}_{\parallel}+\widetilde{\boldsymbol{A}}_{\perp}$. Note that

$$\iiint_{-\infty}^{\infty}f(\boldsymbol{r},t)\boldsymbol{V}(\boldsymbol{r},t)\mathrm{d}\boldsymbol{r}=(2\pi)^{-3}\iiint_{-\infty}^{\infty}\tilde{f}(\boldsymbol{k},t)\widetilde{\boldsymbol{V}}(-\boldsymbol{k},t)\mathrm{d}\boldsymbol{k}.\qquad(7)$$

This is because $\iiint_{-\infty}^{\infty}\exp[\mathrm{i}(\boldsymbol{k}+\boldsymbol{k}')\cdot\boldsymbol{r}]\,\mathrm{d}\boldsymbol{r}=(2\pi)^3\delta(\boldsymbol{k}+\boldsymbol{k}')$. For real-valued $\boldsymbol{V}(\boldsymbol{r},t)$, we have $\widetilde{\boldsymbol{V}}(-\boldsymbol{k},t)=\widetilde{\boldsymbol{V}}^*(\boldsymbol{k},t)$. Let us also mention in passing that the canonical momentum of a point-particle of mass $m$, charge $q$, and position $\boldsymbol{r}(t)$ at time $t$ is defined as $\boldsymbol{\wp}(t)=m\dot{\boldsymbol{r}}(t)+q\boldsymbol{A}[\boldsymbol{r}(t),t]$.[1,4]

---

**Digression**. Alternatively, the integral in Eq.(6) may be evaluated in the Fourier domain, as follows:

$$\boldsymbol{p}_{\parallel}^{(\text{EM})}(t)=\varepsilon_0\iiint_{-\infty}^{\infty}\boldsymbol{E}_{\parallel}(\boldsymbol{r},t)\times\boldsymbol{B}(\boldsymbol{r},t)\mathrm{d}\boldsymbol{r}$$

$$=(2\pi)^{-6}\iiint_{-\infty}^{\infty}\varepsilon_0\widetilde{\boldsymbol{E}}_{\parallel}(\boldsymbol{k},t)\times\widetilde{\boldsymbol{B}}(\boldsymbol{k}',t)e^{\mathrm{i}(\boldsymbol{k}+\boldsymbol{k}')\cdot\boldsymbol{r}}\mathrm{d}\boldsymbol{k}\mathrm{d}\boldsymbol{k}'\mathrm{d}\boldsymbol{r}$$

$$=(2\pi)^{-6}\iiint_{-\infty}^{\infty}\big[-\mathrm{i}(\boldsymbol{k}/k^2)\textstyle\sum_n q_n e^{-\mathrm{i}\boldsymbol{k}\cdot\boldsymbol{r}_n(t)}\big]\times\big[\mathrm{i}\boldsymbol{k}'\times\widetilde{\boldsymbol{A}}(\boldsymbol{k}',t)\big]e^{\mathrm{i}(\boldsymbol{k}+\boldsymbol{k}')\cdot\boldsymbol{r}}\mathrm{d}\boldsymbol{r}\mathrm{d}\boldsymbol{k}'\mathrm{d}\boldsymbol{k}$$

$$=(2\pi)^{-3}\textstyle\sum_n q_n\iiint_{-\infty}^{\infty}(\boldsymbol{k}/k^2)e^{-\mathrm{i}\boldsymbol{k}\cdot\boldsymbol{r}_n(t)}\times[\boldsymbol{k}'\times\widetilde{\boldsymbol{A}}(\boldsymbol{k}',t)]\delta(\boldsymbol{k}+\boldsymbol{k}')\mathrm{d}\boldsymbol{k}'\mathrm{d}\boldsymbol{k}$$

$$=(2\pi)^{-3}\textstyle\sum_n q_n\iiint_{-\infty}^{\infty}(\boldsymbol{k}/k^2)e^{-\mathrm{i}\boldsymbol{k}\cdot\boldsymbol{r}_n(t)}\times[-\boldsymbol{k}\times\widetilde{\boldsymbol{A}}(-\boldsymbol{k},t)]\mathrm{d}\boldsymbol{k}$$

$$=(2\pi)^{-3}\textstyle\sum_n q_n\iiint_{-\infty}^{\infty}(-\boldsymbol{k}/k^2)e^{\mathrm{i}\boldsymbol{k}\cdot\boldsymbol{r}_n(t)}\times\big[\boldsymbol{k}\times\widetilde{\boldsymbol{A}}(\boldsymbol{k},t)\big]\mathrm{d}\boldsymbol{k}\quad\leftarrow\boxed{\boldsymbol{a}\times(\boldsymbol{b}\times\boldsymbol{c})=(\boldsymbol{a}\cdot\boldsymbol{c})\boldsymbol{b}-(\boldsymbol{a}\cdot\boldsymbol{b})\boldsymbol{c}}$$

$$=(2\pi)^{-3}\textstyle\sum_n q_n\iiint_{-\infty}^{\infty}\{\widetilde{\boldsymbol{A}}(\boldsymbol{k},t)-[\hat{\boldsymbol{\kappa}}\cdot\widetilde{\boldsymbol{A}}(\boldsymbol{k},t)]\hat{\boldsymbol{\kappa}}\}\,e^{\mathrm{i}\boldsymbol{k}\cdot\boldsymbol{r}_n(t)}\mathrm{d}\boldsymbol{k}$$

$$=(2\pi)^{-3}\textstyle\sum_n q_n\iiint_{-\infty}^{\infty}\widetilde{\boldsymbol{A}}_{\perp}(\boldsymbol{k},t)e^{\mathrm{i}\boldsymbol{k}\cdot\boldsymbol{r}_n(t)}\mathrm{d}\boldsymbol{k}=\sum_n q_n\boldsymbol{A}_{\perp}[\boldsymbol{r}_n(t),t].\qquad(8)$$

---

**5. Contribution of the longitudinal *E*-field to electromagnetic angular momentum**. The angular momentum associated with the longitudinal component of the $E$-field is computed by first transforming the $\boldsymbol{E}_{\parallel}$ and $\boldsymbol{B}$ fields to the Fourier domain, as follows:

$$\boldsymbol{L}_{\parallel}^{(\text{EM})}(t)=\iiint_{-\infty}^{\infty}\boldsymbol{r}\times[\varepsilon_0\boldsymbol{E}_{\parallel}(\boldsymbol{r},t)\times\boldsymbol{B}(\boldsymbol{r},t)]\mathrm{d}\boldsymbol{r}$$

$$=(2\pi)^{-6}\iiint_{-\infty}^{\infty}\boldsymbol{r}\times[\varepsilon_0\widetilde{\boldsymbol{E}}_{\parallel}(\boldsymbol{k},t)\times\widetilde{\boldsymbol{B}}(\boldsymbol{k}',t)]e^{\mathrm{i}(\boldsymbol{k}+\boldsymbol{k}')\cdot\boldsymbol{r}}\mathrm{d}\boldsymbol{k}\mathrm{d}\boldsymbol{k}'\mathrm{d}\boldsymbol{r}$$

$$= (2\pi)^{-3} \iiint_{-\infty}^{\infty} [-\mathrm{i}\boldsymbol{\nabla}_k \delta(\boldsymbol{k}+\boldsymbol{k}')] \times [\varepsilon_0 \widetilde{\boldsymbol{E}}_{\parallel}(\boldsymbol{k},t) \times \widetilde{\boldsymbol{B}}(\boldsymbol{k}',t)]\mathrm{d}\boldsymbol{k}\mathrm{d}\boldsymbol{k}'$$

$$\boldsymbol{\nabla}_k \delta(\boldsymbol{k}+\boldsymbol{k}') = \iiint_{-\infty}^{\infty} \mathrm{i}\boldsymbol{r}\, e^{\mathrm{i}(\boldsymbol{k}+\boldsymbol{k}')\cdot\boldsymbol{r}} \mathrm{d}\boldsymbol{r} = \delta'(k_x+k_x')\delta(k_y+k_y')\delta(k_z+k_z')\hat{\boldsymbol{x}} + \delta(k_x+k_x')\delta'(k_y+k_y')\delta(k_z+k_z')\hat{\boldsymbol{y}} + \delta(k_x+k_x')\delta(k_y+k_y')\delta'(k_z+k_z')\hat{\boldsymbol{z}}$$

$$\boldsymbol{\nabla}\times(\psi\boldsymbol{V}) = \psi\boldsymbol{\nabla}\times\boldsymbol{V} + (\boldsymbol{\nabla}\psi)\times\boldsymbol{V}$$

$$= -\mathrm{i}(2\pi)^{-3}\Big\{\iiint_{-\infty}^{\infty} \boldsymbol{\nabla}_k \times [\delta(\boldsymbol{k}+\boldsymbol{k}')\varepsilon_0 \widetilde{\boldsymbol{E}}_{\parallel}(\boldsymbol{k},t) \times \widetilde{\boldsymbol{B}}(\boldsymbol{k}',t)]\mathrm{d}\boldsymbol{k}\mathrm{d}\boldsymbol{k}' \rightarrow 0$$

$$- \iiint_{-\infty}^{\infty} \delta(\boldsymbol{k}+\boldsymbol{k}')\boldsymbol{\nabla}_k \times [\varepsilon_0 \widetilde{\boldsymbol{E}}_{\parallel}(\boldsymbol{k},t) \times \widetilde{\boldsymbol{B}}(\boldsymbol{k}',t)]\mathrm{d}\boldsymbol{k}\mathrm{d}\boldsymbol{k}'\Big\}$$

integrate $\boldsymbol{\nabla}_k \times (\cdots)$ over $\boldsymbol{k}$, then set the intgrand to zero at $\pm\infty$

$$\boldsymbol{\nabla}_k \times [\boldsymbol{U}(\boldsymbol{k}) \times \boldsymbol{V}(\boldsymbol{k}')] = (\boldsymbol{V}\cdot\boldsymbol{\nabla}_k)\boldsymbol{U} - (\boldsymbol{\nabla}_k\cdot\boldsymbol{U})\boldsymbol{V}$$

$$= \mathrm{i}(2\pi)^{-3} \iiint_{-\infty}^{\infty} \delta(\boldsymbol{k}+\boldsymbol{k}')\{[\widetilde{\boldsymbol{B}}(\boldsymbol{k}',t)\cdot\boldsymbol{\nabla}_k]\varepsilon_0\widetilde{\boldsymbol{E}}_{\parallel}(\boldsymbol{k},t) - [\boldsymbol{\nabla}_k\cdot\varepsilon_0\widetilde{\boldsymbol{E}}_{\parallel}(\boldsymbol{k},t)]\widetilde{\boldsymbol{B}}(\boldsymbol{k}',t)\}\mathrm{d}\boldsymbol{k}\mathrm{d}\boldsymbol{k}'. \qquad (9)$$

The two terms appearing in the integrand in Eq.(9) that contain the $\boldsymbol{\nabla}_k$ operator are evaluated by differentiation with respect to $k_x, k_y, k_z$, as elaborated below.

$$[\widetilde{\boldsymbol{B}}(\boldsymbol{k}',t)\cdot\boldsymbol{\nabla}_k]\varepsilon_0\widetilde{\boldsymbol{E}}_{\parallel}(\boldsymbol{k},t) = [\widetilde{\boldsymbol{B}}(\boldsymbol{k}',t)\cdot\boldsymbol{\nabla}_k](-\mathrm{i}\boldsymbol{k}/k^2)\textstyle\sum_n q_n e^{-\mathrm{i}\boldsymbol{k}\cdot\boldsymbol{r}_n(t)}$$

$$= -\mathrm{i}\textstyle\sum_n q_n [\widetilde{\boldsymbol{B}}(\boldsymbol{k}',t)\cdot\boldsymbol{\nabla}_k](\boldsymbol{k}/k^2)e^{-\mathrm{i}\boldsymbol{k}\cdot\boldsymbol{r}_n(t)} \leftarrow (k_x\hat{\boldsymbol{x}}+k_y\hat{\boldsymbol{y}}+k_z\hat{\boldsymbol{z}})e^{-\mathrm{i}[k_x x_n(t)+k_y y_n(t)+k_z z_n(t)]}/(k_x^2+k_y^2+k_z^2)$$

$$= -\mathrm{i}\textstyle\sum_n q_n\{\widetilde{B}_x(\boldsymbol{k}',t)[\hat{\boldsymbol{x}} - \mathrm{i}x_n(t)\boldsymbol{k} - (2k_x\boldsymbol{k}/k^2)] + \widetilde{B}_y(\boldsymbol{k}',t)[\hat{\boldsymbol{y}} - \mathrm{i}y_n(t)\boldsymbol{k} - (2k_y\boldsymbol{k}/k^2)]$$

$$+\widetilde{B}_z(\boldsymbol{k}',t)[\hat{\boldsymbol{z}} - \mathrm{i}z_n(t)\boldsymbol{k} - (2k_z\boldsymbol{k}/k^2)]\}\, e^{-\mathrm{i}\boldsymbol{k}\cdot\boldsymbol{r}_n(t)}/k^2$$

$$= -\mathrm{i}\textstyle\sum_n q_n\{\widetilde{\boldsymbol{B}}(\boldsymbol{k}',t) - \mathrm{i}[\widetilde{\boldsymbol{B}}(\boldsymbol{k}',t)\cdot\boldsymbol{r}_n(t)]\boldsymbol{k} - 2[\boldsymbol{k}\cdot\widetilde{\boldsymbol{B}}(\boldsymbol{k}',t)](\boldsymbol{k}/k^2)\}\, e^{-\mathrm{i}\boldsymbol{k}\cdot\boldsymbol{r}_n(t)}/k^2. \qquad (10)$$

$$\boldsymbol{\nabla}_k\cdot\varepsilon_0\widetilde{\boldsymbol{E}}_{\parallel}(\boldsymbol{k},t) = \boldsymbol{\nabla}_k\cdot[(-\mathrm{i}\boldsymbol{k}/k^2)\textstyle\sum_n q_n e^{-\mathrm{i}\boldsymbol{k}\cdot\boldsymbol{r}_n(t)}] = -\mathrm{i}\sum_n q_n\boldsymbol{\nabla}_k\cdot[(\boldsymbol{k}/k^2)e^{-\mathrm{i}\boldsymbol{k}\cdot\boldsymbol{r}_n(t)}]$$

$$= -\mathrm{i}\textstyle\sum_n q_n[1 - \mathrm{i}\boldsymbol{k}\cdot\boldsymbol{r}_n(t)]\, e^{-\mathrm{i}\boldsymbol{k}\cdot\boldsymbol{r}_n(t)}/k^2. \leftarrow 1 = 3 - 2(k_x^2+k_y^2+k_z^2)/k^2 \qquad (11)$$

Substitution from Eqs.(10) and (11) into Eq.(9), followed by integration over $\boldsymbol{k}$ (while invoking the sifting property of the $\delta$-function) now yields

$$\boldsymbol{\nabla}\cdot\boldsymbol{B}(\boldsymbol{r},t) = 0$$

$$\boldsymbol{L}_{\parallel}^{(\mathrm{EM})}(t) = (2\pi)^{-3}\textstyle\sum_n q_n \iiint_{-\infty}^{\infty}\{\widetilde{\boldsymbol{B}}(\boldsymbol{k}',t) + \mathrm{i}[\widetilde{\boldsymbol{B}}(\boldsymbol{k}',t)\cdot\boldsymbol{r}_n(t)]\boldsymbol{k}' - 2[\boldsymbol{k}'\cdot\widetilde{\boldsymbol{B}}(\boldsymbol{k}',t)](\boldsymbol{k}'/k'^2) \rightarrow 0$$

$$-\widetilde{\boldsymbol{B}}(\boldsymbol{k}',t) - \mathrm{i}[\boldsymbol{k}'\cdot\boldsymbol{r}_n(t)]\widetilde{\boldsymbol{B}}(\boldsymbol{k}',t)\}[e^{\mathrm{i}\boldsymbol{k}'\cdot\boldsymbol{r}_n(t)}/k'^2]\mathrm{d}\boldsymbol{k}'$$

$$= \mathrm{i}(2\pi)^{-3}\textstyle\sum_n q_n \iiint_{-\infty}^{\infty} \boldsymbol{r}_n(t)\times[\boldsymbol{k}'\times\widetilde{\boldsymbol{B}}(\boldsymbol{k}',t)][e^{\mathrm{i}\boldsymbol{k}'\cdot\boldsymbol{r}_n(t)}/k'^2]\mathrm{d}\boldsymbol{k}' \leftarrow \boldsymbol{B}(\boldsymbol{r},t) = \boldsymbol{\nabla}\times\boldsymbol{A}(\boldsymbol{r},t)$$

$$= \mathrm{i}(2\pi)^{-3}\textstyle\sum_n q_n\boldsymbol{r}_n(t)\times\iiint_{-\infty}^{\infty}\boldsymbol{k}'\times[\mathrm{i}\boldsymbol{k}'\times\widetilde{\boldsymbol{A}}(\boldsymbol{k}',t)][e^{\mathrm{i}\boldsymbol{k}'\cdot\boldsymbol{r}_n(t)}/k'^2]\mathrm{d}\boldsymbol{k}'$$

$$= (2\pi)^{-3}\textstyle\sum_n q_n\boldsymbol{r}_n(t)\times\iiint_{-\infty}^{\infty}\widetilde{\boldsymbol{A}}_{\perp}(\boldsymbol{k}',t)e^{\mathrm{i}\boldsymbol{k}'\cdot\boldsymbol{r}_n(t)}\mathrm{d}\boldsymbol{k}'$$

$$= \textstyle\sum_n q_n\boldsymbol{r}_n(t)\times\boldsymbol{A}_{\perp}[\boldsymbol{r}_n(t),t]. \qquad (12)$$

The angular momentum associated with $\boldsymbol{E}_{\parallel}$ is thus seen to be localized at the positions $\boldsymbol{r}_n(t)$ of the point-charges, given by the (simultaneous) sum over all the charges of the cross-product of $\boldsymbol{r}_n(t)$ and the EM linear momentum $q_n\boldsymbol{A}_{\perp}[\boldsymbol{r}_n(t),t]$ assigned to each point-particle.

**6. Maxwell's equations in the presence of point particles**. The vector potential $\boldsymbol{A}(\boldsymbol{r},t)$, defined as the field satisfying $\boldsymbol{\nabla}\times\boldsymbol{A}=\boldsymbol{B}$, ensures the satisfaction of Maxwell's 4th equation, $\boldsymbol{\nabla}\cdot\boldsymbol{B}=0$.[2,3] Similarly, the scalar potential $\psi(\boldsymbol{r},t)$, defined such that $\boldsymbol{E}=-\boldsymbol{\nabla}\psi-\partial\boldsymbol{A}/\partial t$, guarantees that Maxwell's 3rd equation, $\boldsymbol{\nabla}\times\boldsymbol{E}=-\partial\boldsymbol{B}/\partial t$, is always satisfied.[2,3] The longitudinal part of the $E$-field is directly obtained from Maxwell's 1st equation, $\boldsymbol{\nabla}\cdot(\varepsilon_0\boldsymbol{E})=\rho_{\text{free}}$, as

$$\boldsymbol{E}_{\parallel}(\boldsymbol{r},t)=\frac{1}{4\pi\varepsilon_0}\sum_n q_n[\boldsymbol{r}-\boldsymbol{r}_n(t)]/|\boldsymbol{r}-\boldsymbol{r}_n(t)|^3. \tag{13}$$

In the Coulomb gauge, where $\boldsymbol{A}_{\parallel}(\boldsymbol{r},t)=0$, we have $\boldsymbol{E}_{\parallel}=-\boldsymbol{\nabla}\psi$, which leads to $\psi(\boldsymbol{r},t)=\sum_n q_n/[4\pi\varepsilon_0|\boldsymbol{r}-\boldsymbol{r}_n(t)|]$. The remaining Maxwell equation, $\boldsymbol{\nabla}\times\boldsymbol{B}=\mu_0\boldsymbol{J}_{\text{free}}+\mu_0\varepsilon_0\partial\boldsymbol{E}/\partial t$,[§] now yields a relation between $\boldsymbol{A}_{\perp}$ and the transverse component $\boldsymbol{J}_{\perp(\text{free})}$ of $\boldsymbol{J}_{\text{free}}(\boldsymbol{r},t)$, as follows:[2,3]

$$c^2\boldsymbol{\nabla}\times(\boldsymbol{\nabla}\times\boldsymbol{A}_{\perp})+\partial^2\boldsymbol{A}_{\perp}/\partial t^2=\boldsymbol{J}_{\perp(\text{free})}/\varepsilon_0. \tag{14}$$

← $c=(\mu_0\varepsilon_0)^{-½}$ is the speed of light in vacuum.

The longitudinal component $\boldsymbol{J}_{\parallel(\text{free})}+\partial(\varepsilon_0\boldsymbol{E}_{\parallel})/\partial t=0$ (i.e., the remaining part) of Maxwell's 2nd equation is now seen to be automatically satisfied in light of the charge-current continuity equation, $\boldsymbol{\nabla}\cdot\boldsymbol{J}_{\text{free}}+\partial\rho_{\text{free}}/\partial t=0$. An alternative way of verifying this assertion is by Fourier transforming $\boldsymbol{J}_{\text{free}}(\boldsymbol{r},t)=\sum_n q_n\dot{\boldsymbol{r}}_n(t)\delta[\boldsymbol{r}-\boldsymbol{r}_n(t)]$ to obtain $\tilde{\boldsymbol{J}}_{\text{free}}(\boldsymbol{k},t)=\sum_n q_n\dot{\boldsymbol{r}}_n(t)e^{-\mathrm{i}\boldsymbol{k}\cdot\boldsymbol{r}_n(t)}$, which leads to

$$\tilde{\boldsymbol{J}}_{\parallel(\text{free})}(\boldsymbol{k},t)=(\boldsymbol{k}/k^2)\sum_n q_n\boldsymbol{k}\cdot\dot{\boldsymbol{r}}_n(t)e^{-\mathrm{i}\boldsymbol{k}\cdot\boldsymbol{r}_n(t)}. \tag{15}$$

Given that $\varepsilon_0\tilde{\boldsymbol{E}}_{\parallel}(\boldsymbol{k},t)=-\mathrm{i}(\boldsymbol{k}/k^2)\tilde{\rho}_{\text{free}}(\boldsymbol{k},t)=-\mathrm{i}(\boldsymbol{k}/k^2)\sum_n q_n e^{-\mathrm{i}\boldsymbol{k}\cdot\boldsymbol{r}_n(t)}$, it is readily seen that

$$\partial(\varepsilon_0\tilde{\boldsymbol{E}}_{\parallel})/\partial t=-(\boldsymbol{k}/k^2)\sum_n q_n\boldsymbol{k}\cdot\dot{\boldsymbol{r}}_n(t)e^{-\mathrm{i}\boldsymbol{k}\cdot\boldsymbol{r}_n(t)}. \tag{16}$$

We have thus arrived at $\tilde{\boldsymbol{J}}_{\parallel(\text{free})}+\partial(\varepsilon_0\tilde{\boldsymbol{E}}_{\parallel})/\partial t=0$, which is the Fourier domain version of the desired identity.

**7. Electromagnetic radiation and normal variables**. The transverse fields $\tilde{\boldsymbol{E}}_{\perp}(\boldsymbol{k},t)$, $\tilde{\boldsymbol{B}}(\boldsymbol{k},t)$, $\tilde{\boldsymbol{A}}_{\perp}(\boldsymbol{k},t)$, and $\tilde{\boldsymbol{J}}_{\perp}(\boldsymbol{k},t)$ contain the independent dynamic variables of the EM radiation. For each $k$-vector $\boldsymbol{k}=k\hat{\boldsymbol{\kappa}}$, one can define a pair of (generally complex[**]) orthogonal unit-vectors $\hat{\boldsymbol{e}}_1$ and $\hat{\boldsymbol{e}}_2$, both perpendicular to $\hat{\boldsymbol{\kappa}}$, so that two independent (i.e., orthogonally polarized) radiative modes are associated with $(\boldsymbol{k},\hat{\boldsymbol{e}}_1)$ and $(\boldsymbol{k},\hat{\boldsymbol{e}}_2)$. Each of the two orthogonally-polarized Fourier components possessing the wave-vector $\boldsymbol{k}$ can then be identified with their scalar field amplitudes $\tilde{A}_{\perp,\ell}(t), \tilde{E}_{\perp,\ell}(t), \tilde{B}_{\ell}(t)$, where the index $\ell$ specifies not only the discretized $k$-vector,

$$\boldsymbol{k}_{\ell}=(2\pi/L)(n_x\hat{\boldsymbol{x}}+n_y\hat{\boldsymbol{y}}+n_z\hat{\boldsymbol{z}}), \tag{17}$$

but also a corresponding state of polarization (i.e., $\hat{\boldsymbol{e}}_1$ or $\hat{\boldsymbol{e}}_2$). The mode index may thus be written as $\ell=(n_x,n_y,n_z,s)$, with $s=1$ if the polarization vector is $\hat{\boldsymbol{e}}_{\ell}=\hat{\boldsymbol{e}}_1$, and $s=2$ if $\hat{\boldsymbol{e}}_{\ell}=\hat{\boldsymbol{e}}_2$. Here, $L^3$ is the volume of the (very large) cube to which the various modes of radiation are assumed to be confined, while $(n_x, n_y, n_z)$ are the positive, zero, and negative integers that uniquely identify each individual $k$-vector. Whereas the polarization vector assigned to $\tilde{E}_{\perp,\ell}(t)$ and $\tilde{A}_{\perp,\ell}(t)$ is $\hat{\boldsymbol{e}}_\ell$, the corresponding unit-vector for $\tilde{B}_\ell(t)$ is $\hat{\boldsymbol{\kappa}}_\ell \times \hat{\boldsymbol{e}}_\ell$. The discretized field amplitudes and their driving source term $\tilde{J}_{\perp,\ell}(t)$ satisfy the following relations:

$$\tilde{E}_{\perp,\ell}(t) = -\frac{\mathrm{d}\tilde{A}_{\perp,\ell}(t)}{\mathrm{d}t}. \tag{18}$$

$$-\mathrm{i}k_\ell \tilde{B}_\ell(t) = \mu_0 \tilde{J}_{\perp,\ell}(t) + \mu_0\varepsilon_0 \frac{\mathrm{d}\tilde{E}_{\perp,\ell}(t)}{\mathrm{d}t} \quad \rightarrow \quad c^2 k_\ell^2 \tilde{A}_{\perp,\ell}(t) = \varepsilon_0^{-1}\tilde{J}_{\perp,\ell}(t) + \frac{\mathrm{d}\tilde{E}_{\perp,\ell}(t)}{\mathrm{d}t}. \tag{19}$$

---

[§] The vacuum permittivity and permeability are $\varepsilon_0=8.854\cdots\times10^{-12}$ [farad/m] and $\mu_0=4\pi\times10^{-7}$ [henry/m].

[**] The complex unit-vectors $\hat{\boldsymbol{e}}_1$ and $\hat{\boldsymbol{e}}_2$ satisfy the relations $\hat{\boldsymbol{e}}_1\cdot\hat{\boldsymbol{e}}_1^*=\hat{\boldsymbol{e}}_2\cdot\hat{\boldsymbol{e}}_2^*=1$, and $\hat{\boldsymbol{e}}_1\cdot\hat{\boldsymbol{e}}_2^*=0$. Also, both $\hat{\boldsymbol{e}}_1$ and $\hat{\boldsymbol{e}}_2$ are orthogonal to $\hat{\boldsymbol{\kappa}}$; that is, $\hat{\boldsymbol{\kappa}}\cdot\hat{\boldsymbol{e}}_1=\hat{\boldsymbol{\kappa}}\cdot\hat{\boldsymbol{e}}_2=0$. With rare exceptions, $\hat{\boldsymbol{\kappa}}\times\hat{\boldsymbol{e}}_1\neq\hat{\boldsymbol{e}}_2$ and, similarly, $\hat{\boldsymbol{\kappa}}\times\hat{\boldsymbol{e}}_2\neq\hat{\boldsymbol{e}}_1$, because $(\hat{\boldsymbol{\kappa}}\times\hat{\boldsymbol{e}}_1)\cdot\hat{\boldsymbol{e}}_1^*=(\hat{\boldsymbol{e}}_1\times\hat{\boldsymbol{e}}_1^*)\cdot\hat{\boldsymbol{\kappa}}=[(\boldsymbol{e}_1'+\mathrm{i}\boldsymbol{e}_1'')\times(\boldsymbol{e}_1'-\mathrm{i}\boldsymbol{e}_1'')]\cdot\hat{\boldsymbol{\kappa}}=2\mathrm{i}(\boldsymbol{e}_1''\times\boldsymbol{e}_1')\cdot\hat{\boldsymbol{\kappa}}\neq0$. (The aforementioned rare exceptions arise when $\boldsymbol{e}_1'$ is aligned with $\boldsymbol{e}_1''$, or when $\hat{\boldsymbol{e}}_1$ and $\hat{\boldsymbol{e}}_2$ are real-valued.) Note that, if $\hat{\boldsymbol{e}}_1$ happens to represent a right-elliptical state of polarization, then $\hat{\boldsymbol{e}}_2$ corresponds to a left-elliptical state, whereas $\hat{\boldsymbol{\kappa}}\times\hat{\boldsymbol{e}}_1$ continues to represent a right-elliptical state.

These two coupled equations in $\tilde{A}_{\perp,\ell}$ and $\tilde{E}_{\perp,\ell}$ can be decoupled by introducing the so-called *normal variables* $\alpha_\ell(t) = \zeta_\ell(\omega_\ell \tilde{A}_{\perp,\ell} - \mathrm{i}\tilde{E}_{\perp,\ell})$ and $\beta_\ell(t) = \zeta_\ell(\omega_\ell \tilde{A}_{\perp,\ell} + \mathrm{i}\tilde{E}_{\perp,\ell})$, with $\zeta_\ell$ being a real-valued scale parameter (to be fixed later), and $\omega_\ell = ck_\ell$ being a non-negative constant.[1,4] Subsequently, Eqs.(18) and (19) are rearranged to yield the following (decoupled) equations:

$$\frac{\mathrm{d}\alpha_\ell(t)}{\mathrm{d}t} + \mathrm{i}\omega_\ell \alpha_\ell(t) = \mathrm{i}(\zeta_\ell/\varepsilon_0)\tilde{J}_{\perp,\ell}(t). \tag{20}$$

$$\frac{\mathrm{d}\beta_\ell(t)}{\mathrm{d}t} - \mathrm{i}\omega_\ell \beta_\ell(t) = -\mathrm{i}(\zeta_\ell/\varepsilon_0)\tilde{J}_{\perp,\ell}(t). \tag{21}$$

In terms of the solutions $\alpha_\ell(t)$ and $\beta_\ell(t)$ of the above (decoupled) equations, we will have $\tilde{A}_{\perp,\ell} = (\alpha_\ell + \beta_\ell)/(2\zeta_\ell\omega_\ell)$ and $\tilde{E}_{\perp,\ell} = \mathrm{i}(\alpha_\ell - \beta_\ell)/(2\zeta_\ell)$. The overall solutions $\boldsymbol{A}_\perp$, $\boldsymbol{E}_\perp$, and $\boldsymbol{B}$ of Maxwell's equations may now be written as[††]

$V = L^3$ is the volume occupied by each mode in the $xyz$ space.

$$\boldsymbol{A}_\perp(\boldsymbol{r}, t) = L^{-3}\textstyle\sum_\ell \tilde{A}_{\perp,\ell}(t) e^{\mathrm{i}\boldsymbol{k}_\ell\cdot\boldsymbol{r}}\hat{\boldsymbol{e}}_\ell = \sum_\ell (2V\zeta_\ell\omega_\ell)^{-1}[\alpha_\ell(t) + \beta_\ell(t)] e^{\mathrm{i}\boldsymbol{k}_\ell\cdot\boldsymbol{r}}\hat{\boldsymbol{e}}_\ell. \tag{22}$$

$$\boldsymbol{E}_\perp(\boldsymbol{r}, t) = L^{-3}\textstyle\sum_\ell \tilde{E}_{\perp,\ell}(t) e^{\mathrm{i}\boldsymbol{k}_\ell\cdot\boldsymbol{r}}\hat{\boldsymbol{e}}_\ell = \mathrm{i}\sum_\ell (2V\zeta_\ell)^{-1}[\alpha_\ell(t) - \beta_\ell(t)] e^{\mathrm{i}\boldsymbol{k}_\ell\cdot\boldsymbol{r}}\hat{\boldsymbol{e}}_\ell. \tag{23}$$

$$\boldsymbol{B}(\boldsymbol{r}, t) = L^{-3}\textstyle\sum_\ell \tilde{A}_{\perp,\ell}(t) e^{\mathrm{i}\boldsymbol{k}_\ell\cdot\boldsymbol{r}}(\mathrm{i}\boldsymbol{k}_\ell \times \hat{\boldsymbol{e}}_\ell) = \mathrm{i}\sum_\ell (2V\zeta_\ell\omega_\ell)^{-1}[\alpha_\ell(t) + \beta_\ell(t)] e^{\mathrm{i}\boldsymbol{k}_\ell\cdot\boldsymbol{r}}(\boldsymbol{k}_\ell \times \hat{\boldsymbol{e}}_\ell). \tag{24}$$

Given that the $\boldsymbol{A}_\perp$, $\boldsymbol{E}_\perp$, and $\boldsymbol{B}$ fields are real-valued functions of $(\boldsymbol{r}, t)$, their Fourier components automatically satisfy the constraints $\tilde{\boldsymbol{A}}_\perp(-\boldsymbol{k}, t) = \tilde{\boldsymbol{A}}_\perp^*(\boldsymbol{k}, t)$, $\tilde{\boldsymbol{E}}_\perp(-\boldsymbol{k}, t) = \tilde{\boldsymbol{E}}_\perp^*(\boldsymbol{k}, t)$, and $\tilde{\boldsymbol{B}}(-\boldsymbol{k}, t) = \tilde{\boldsymbol{B}}^*(\boldsymbol{k}, t)$. Consequently, $\hat{\boldsymbol{e}}_{-\ell} = \hat{\boldsymbol{e}}_\ell^*$, $\tilde{A}_{\perp,-\ell} = \tilde{A}_{\perp,\ell}^*$, and $\tilde{E}_{\perp,-\ell} = \tilde{E}_{\perp,\ell}^*$; here the subscript $-\ell$ refers to Fourier components whose $k$-vectors are flipped to $-\boldsymbol{k}$ while retaining their polarization state—aside from the necessary conversions $\hat{\boldsymbol{e}}_1 \rightarrow \hat{\boldsymbol{e}}_1^*$ and $\hat{\boldsymbol{e}}_2 \rightarrow \hat{\boldsymbol{e}}_2^*$. The normal variables are thus seen to satisfy the constraints $\alpha_{-\ell}(t) = \beta_\ell^*(t)$ and $\beta_{-\ell}(t) = \alpha_\ell^*(t)$.

In Eqs.(22)-(24), we may now separate the sum over $\ell$ into one part that contains only the normal variables $\alpha_\ell(t)$, and a second part containing only $\beta_\ell(t)$. The second sum would not change if we switched the index $\ell$ to $-\ell$, because, either way, the sum includes all possible values of $\ell$. Upon substituting $-\boldsymbol{k}_\ell$ for $\boldsymbol{k}_\ell$ in the second sum, then replacing $\hat{\boldsymbol{e}}_{-\ell}$ with $\hat{\boldsymbol{e}}_\ell^*$, and $\beta_{-\ell}(t)$ with $\alpha_\ell^*(t)$, we arrive at

$$\boldsymbol{A}_\perp(\boldsymbol{r}, t) = \textstyle\sum_\ell (2V\zeta_\ell\omega_\ell)^{-1}\left[\alpha_\ell(t) e^{\mathrm{i}\boldsymbol{k}_\ell\cdot\boldsymbol{r}}\hat{\boldsymbol{e}}_\ell + \alpha_\ell^*(t) e^{-\mathrm{i}\boldsymbol{k}_\ell\cdot\boldsymbol{r}}\hat{\boldsymbol{e}}_\ell^*\right]. \tag{25}$$

---

[††] For any field $\boldsymbol{F}(\boldsymbol{r}, t)$ residing in the 3D space $\boldsymbol{r}$, the Fourier integral is $\tilde{\boldsymbol{F}}(\boldsymbol{k}, t) = \iiint_{-\infty}^{\infty} \boldsymbol{F}(\boldsymbol{r}, t) e^{-\mathrm{i}\boldsymbol{k}\cdot\boldsymbol{r}}\mathrm{d}\boldsymbol{r}$, while the inverse Fourier integral is $\boldsymbol{F}(\boldsymbol{r}, t) = (2\pi)^{-3}\iiint_{-\infty}^{\infty}\tilde{\boldsymbol{F}}(\boldsymbol{k}, t) e^{\mathrm{i}\boldsymbol{k}\cdot\boldsymbol{r}}\mathrm{d}\boldsymbol{k}$. Confining the field to a large $L \times L \times L$ cube in the $xyz$ domain and imposing periodic boundary conditions on $\boldsymbol{F}(\boldsymbol{r}, t)$ results in a discretization of the $k$-space, where discrete $k$-vectors $\boldsymbol{k}_\ell = (2\pi/L)(n_x\hat{\boldsymbol{x}} + n_y\hat{\boldsymbol{y}} + n_z\hat{\boldsymbol{z}})$ reside within differential volume elements $\mathrm{d}\boldsymbol{k} = \mathrm{d}k_x\mathrm{d}k_y\mathrm{d}k_z = (2\pi/L)^3$, and the inverse Fourier integral reduces to $\boldsymbol{F}(\boldsymbol{r}, t) = L^{-3}\sum_\ell \tilde{\boldsymbol{F}}(\boldsymbol{k}_\ell, t) e^{\mathrm{i}\boldsymbol{k}_\ell\cdot\boldsymbol{r}}$.

$$\boldsymbol{E}_\perp(\boldsymbol{r},t) = \mathrm{i}\textstyle\sum_\ell (2V\zeta_\ell)^{-1}\big[\alpha_\ell(t)e^{\mathrm{i}\boldsymbol{k}_\ell\cdot\boldsymbol{r}}\hat{\boldsymbol{e}}_\ell - \alpha_\ell^*(t)e^{-\mathrm{i}\boldsymbol{k}_\ell\cdot\boldsymbol{r}}\hat{\boldsymbol{e}}_\ell^*\big]. \tag{26}$$

$$\boldsymbol{B}(\boldsymbol{r},t) = \mathrm{i}\textstyle\sum_\ell (2V\zeta_\ell c)^{-1}\big[\alpha_\ell(t)e^{\mathrm{i}\boldsymbol{k}_\ell\cdot\boldsymbol{r}}(\hat{\boldsymbol{\kappa}}_\ell\times\hat{\boldsymbol{e}}_\ell) - \alpha_\ell^*(t)e^{-\mathrm{i}\boldsymbol{k}_\ell\cdot\boldsymbol{r}}(\hat{\boldsymbol{\kappa}}_\ell\times\hat{\boldsymbol{e}}_\ell^*)\big]. \tag{27}$$

The inclusion of the scale-factor $\zeta_\ell$ in the definition of the normal variables $\alpha_\ell(t)$ and $\beta_\ell(t)$ requires further scrutiny. As revealed by Eqs.(20) and (21), this scale-factor has no influence on the particular solution of each equation, considering that the driving term $\tilde{J}_{\perp,\ell}(t)$ is simultaneously scaled by the same factor. It is only in the context of the homogeneous solutions of Eqs.(20) and (21), namely, $\alpha_\ell(t) = \alpha_{\ell 0}e^{-\mathrm{i}\omega_\ell t}$ and $\beta_\ell(t) = \beta_{\ell 0}e^{\mathrm{i}\omega_\ell t}$, that $\zeta_\ell$ can be called upon to adjust the values of the integration constants $\alpha_{\ell 0}$ and $\beta_{\ell 0}$. We will see in the next section that, by setting $\alpha_{\ell 0} = \beta_{\ell 0} = 1$ and $\zeta_\ell = (\varepsilon_0/2V\hbar\omega_\ell)^{½}$, the energy content of a single-photon-state $|1\rangle$ within the traveling-wave mode $(\omega_\ell, \boldsymbol{k}_\ell, \hat{\boldsymbol{e}}_\ell)$ in free space ends up being equal to $\hbar\omega_\ell$, which is what will be needed when the EM field is quantized. Another noteworthy feature of these homogeneous solutions is that both $\alpha_\ell(t) = e^{-\mathrm{i}\omega_\ell t}$ and $\beta_\ell(t) = e^{\mathrm{i}\omega_\ell t}$ are subtly embedded in Eqs.(25)-(27). This is because the traveling plane-wave modes containing the terms $e^{\pm\mathrm{i}(\boldsymbol{k}_\ell\cdot\boldsymbol{r}-\omega_\ell t)}$ (associated with $\alpha_\ell(t)$) turn into $e^{\mp\mathrm{i}(\boldsymbol{k}_\ell\cdot\boldsymbol{r}+\omega_\ell t)}$ (associated with $\beta_\ell(t)$) when $\boldsymbol{k}_\ell \to \boldsymbol{k}_{-\ell} = -\boldsymbol{k}_\ell$. In other words, for each propagation direction $\hat{\boldsymbol{\kappa}}$ in free space, Eqs.(25)-(27) contain one traveling plane-wave along the direction of $\hat{\boldsymbol{\kappa}}$, and a second (independent) traveling plane-wave along $-\hat{\boldsymbol{\kappa}}$.

**8. Energy and linear momentum of the radiation field**. Upon transforming to the Fourier domain, the EM radiation energy can be expressed as a sum over the product $\alpha_\ell(t)\alpha_\ell^*(t)$ of the normal variables, as follows:

$$\begin{aligned}
\mathcal{E}(t) &= \textstyle\iiint_{-\infty}^{\infty}[½\varepsilon_0\boldsymbol{E}_\perp(\boldsymbol{r},t)\cdot\boldsymbol{E}_\perp(\boldsymbol{r},t) + ½\mu_0^{-1}\boldsymbol{B}(\boldsymbol{r},t)\cdot\boldsymbol{B}(\boldsymbol{r},t)]\mathrm{d}\boldsymbol{r}\\
&= (2\pi)^{-6}\textstyle\iiint_{-\infty}^{\infty}[½\varepsilon_0\tilde{\boldsymbol{E}}_\perp(\boldsymbol{k},t)\cdot\tilde{\boldsymbol{E}}_\perp(\boldsymbol{k}',t) + ½\mu_0^{-1}\tilde{\boldsymbol{B}}(\boldsymbol{k},t)\cdot\tilde{\boldsymbol{B}}(\boldsymbol{k}',t)]e^{\mathrm{i}(\boldsymbol{k}+\boldsymbol{k}')\cdot\boldsymbol{r}}\mathrm{d}\boldsymbol{k}\mathrm{d}\boldsymbol{k}'\mathrm{d}\boldsymbol{r}\\
&= (2\pi)^{-3}\textstyle\iiint_{-\infty}^{\infty}½\varepsilon_0[\tilde{\boldsymbol{E}}_\perp(\boldsymbol{k},t)\cdot\tilde{\boldsymbol{E}}_\perp(\boldsymbol{k}',t) + c^2\tilde{\boldsymbol{B}}(\boldsymbol{k},t)\cdot\tilde{\boldsymbol{B}}(\boldsymbol{k}',t)]\delta(\boldsymbol{k}+\boldsymbol{k}')\mathrm{d}\boldsymbol{k}'\mathrm{d}\boldsymbol{k}\\
&= (2\pi)^{-3}\textstyle\iiint_{-\infty}^{\infty}½\varepsilon_0[\tilde{\boldsymbol{E}}_\perp(\boldsymbol{k},t)\cdot\tilde{\boldsymbol{E}}_\perp(-\boldsymbol{k},t) + c^2\tilde{\boldsymbol{B}}(\boldsymbol{k},t)\cdot\tilde{\boldsymbol{B}}(-\boldsymbol{k},t)]\mathrm{d}\boldsymbol{k}
\end{aligned}$$

$$\boxed{\tilde{\boldsymbol{B}}\cdot\tilde{\boldsymbol{B}}^* = (\mathrm{i}\boldsymbol{k}\times\tilde{\boldsymbol{A}}_\perp)\cdot(-\mathrm{i}\boldsymbol{k}\times\tilde{\boldsymbol{A}}_\perp^*) = [(\boldsymbol{k}\times\tilde{\boldsymbol{A}}_\perp)\times\boldsymbol{k}]\cdot\tilde{\boldsymbol{A}}_\perp^* = k^2\tilde{\boldsymbol{A}}_\perp\cdot\tilde{\boldsymbol{A}}_\perp^*}$$

$$\begin{aligned}
&= (2\pi)^{-3}\textstyle\iiint_{-\infty}^{\infty}½\varepsilon_0[\tilde{\boldsymbol{E}}_\perp(\boldsymbol{k},t)\cdot\tilde{\boldsymbol{E}}_\perp^*(\boldsymbol{k},t) + c^2k^2\tilde{\boldsymbol{A}}_\perp(\boldsymbol{k},t)\cdot\tilde{\boldsymbol{A}}_\perp^*(\boldsymbol{k},t)]\mathrm{d}\boldsymbol{k}\\
&= ½\varepsilon_0L^{-3}\textstyle\sum_\ell(\tilde{E}_{\perp,\ell}\tilde{E}_{\perp,\ell}^* + \omega_\ell^2\tilde{A}_{\perp,\ell}\tilde{A}_{\perp,\ell}^*)
\end{aligned}$$

$$\boxed{\textstyle\sum_\ell\tilde{E}_{\perp,\ell}\tilde{A}_{\perp,\ell}^* = \sum_\ell\tilde{E}_{\perp,-\ell}\tilde{A}_{\perp,-\ell}^* = \sum_\ell\tilde{E}_{\perp,\ell}^*\tilde{A}_{\perp,\ell}}$$

$$\begin{aligned}
&= ½\varepsilon_0L^{-3}\textstyle\sum_\ell[(\omega_\ell\tilde{A}_{\perp,\ell} - \mathrm{i}\tilde{E}_{\perp,\ell})(\omega_\ell\tilde{A}_{\perp,\ell}^* + \mathrm{i}\tilde{E}_{\perp,\ell}^*) + \mathrm{i}\omega_\ell(\cancel{\tilde{E}_{\perp,\ell}\tilde{A}_{\perp,\ell}^*} - \cancel{\tilde{A}_{\perp,\ell}\tilde{E}_{\perp,\ell}^*})]\\
&= [\varepsilon_0/(2V)]\textstyle\sum_\ell\alpha_\ell(t)\alpha_\ell^*(t)/\zeta_\ell^2.
\end{aligned} \tag{28}$$

Similarly, the radiation field's linear momentum can be written as a sum over the product $\alpha_\ell(t)\alpha_\ell^*(t)$ of the normal variables, namely,

$$\begin{aligned}
\boldsymbol{p}^{(\mathrm{EM})}(t) &= \varepsilon_0\textstyle\iiint_{-\infty}^{\infty}\boldsymbol{E}_\perp(\boldsymbol{r},t)\times\boldsymbol{B}(\boldsymbol{r},t)\mathrm{d}\boldsymbol{r}\\
&= (2\pi)^{-6}\varepsilon_0\textstyle\iiint_{-\infty}^{\infty}\tilde{\boldsymbol{E}}_\perp(\boldsymbol{k},t)\times\tilde{\boldsymbol{B}}(\boldsymbol{k}',t)e^{\mathrm{i}(\boldsymbol{k}+\boldsymbol{k}')\cdot\boldsymbol{r}}\mathrm{d}\boldsymbol{k}\mathrm{d}\boldsymbol{k}'\mathrm{d}\boldsymbol{r}\\
&= (2\pi)^{-3}\varepsilon_0\textstyle\iiint_{-\infty}^{\infty}\tilde{\boldsymbol{E}}_\perp(\boldsymbol{k},t)\times[\mathrm{i}\boldsymbol{k}'\times\tilde{\boldsymbol{A}}_\perp(\boldsymbol{k}',t)]\delta(\boldsymbol{k}+\boldsymbol{k}')\mathrm{d}\boldsymbol{k}'\mathrm{d}\boldsymbol{k}
\end{aligned}$$

$$= -\mathrm{i}(2\pi)^{-3}\varepsilon_0 \iiint_{-\infty}^{\infty} \tilde{\boldsymbol{E}}_\perp(\boldsymbol{k},t) \times [\boldsymbol{k} \times \tilde{\boldsymbol{A}}_\perp(-\boldsymbol{k},t)]\mathrm{d}\boldsymbol{k}$$

$$= -\mathrm{i}(2\pi)^{-3}\varepsilon_0 \iiint_{-\infty}^{\infty} [\tilde{\boldsymbol{E}}_\perp(\boldsymbol{k},t) \cdot \tilde{\boldsymbol{A}}_\perp^*(\boldsymbol{k},t)]\boldsymbol{k}\mathrm{d}\boldsymbol{k}$$

$$= -\mathrm{i}\varepsilon_0 L^{-3} \textstyle\sum_\ell [\tilde{E}_{\perp,\ell}(t)\tilde{A}_{\perp,\ell}^*(t)]\boldsymbol{k}_\ell$$

This term is the same for $\ell$ and $-\ell$, but $\boldsymbol{k}_{-\ell} = -\boldsymbol{k}_\ell$

$$= [\varepsilon_0/(2V)] \textstyle\sum_\ell [(\omega_\ell \tilde{A}_{\perp,\ell} - \mathrm{i}\tilde{E}_{\perp,\ell})(\omega_\ell \tilde{A}_{\perp,\ell}^* + \mathrm{i}\tilde{E}_{\perp,\ell}^*) - (\omega_\ell^2 \tilde{A}_{\perp,\ell}\tilde{A}_{\perp,\ell}^* + \tilde{E}_{\perp,\ell}\tilde{E}_{\perp,\ell}^*)]\,\boldsymbol{k}_\ell/\omega_\ell$$

$\sum_\ell \tilde{A}_{\perp,\ell}\tilde{E}_{\perp,\ell}^*\boldsymbol{k}_\ell = \sum_\ell \tilde{A}_{\perp,-\ell}\tilde{E}_{\perp,-\ell}^*\boldsymbol{k}_{-\ell} = -\sum_\ell \tilde{A}_{\perp,\ell}^*\tilde{E}_{\perp,\ell}\boldsymbol{k}_\ell$

$\boldsymbol{k}_\ell = (\omega_\ell/c)\hat{\boldsymbol{\kappa}}_\ell$

$$= [\varepsilon_0/(2Vc)] \textstyle\sum_\ell \alpha_\ell(t)\alpha_\ell^*(t)\hat{\boldsymbol{\kappa}}_\ell/\zeta_\ell^2. \qquad (29)$$

Upon quantization, the normal variable $\alpha_\ell$ becomes the annihilation operator $\hat{a}_\ell$, while its complex conjugate $\alpha_\ell^*$ becomes the creation operator $\hat{a}_\ell^\dagger$. The coefficient $\varepsilon_0/(2V\zeta_\ell^2)$ appearing in both Eqs.(28) and (29) must then be equated with the single-photon energy $\hbar\omega_\ell$. In this way, the scale-factor is found to be $\zeta_\ell = (\varepsilon_0/2V\hbar\omega_\ell)^{½}$.

**9. Spin and orbital angular momenta of the radiation field**. The general expression of the total angular momentum carried by an arbitrary EM wavepacket (finite in its spatial extent) is

$$\boldsymbol{\mathcal{J}}(t) = \iiint_{-\infty}^{\infty} \boldsymbol{r} \times [\varepsilon_0 \boldsymbol{E}(\boldsymbol{r},t) \times \boldsymbol{B}(\boldsymbol{r},t)]\mathrm{d}v. \qquad (30)$$

A first step in evaluating the above integral involves a rewriting of the Poynting vector[2,3] $\boldsymbol{E} \times \boldsymbol{B}/\mu_0$ that appears within the integrand, as follows:

$$\begin{aligned}\boldsymbol{E}(\boldsymbol{r},t) \times \boldsymbol{B}(\boldsymbol{r},t) = \boldsymbol{E} \times (\boldsymbol{\nabla} \times \boldsymbol{A}) &= [E_y(\partial_x A_y - \partial_y A_x) - E_z(\partial_z A_x - \partial_x A_z)]\hat{\boldsymbol{x}} \\ &+ [E_z(\partial_y A_z - \partial_z A_y) - E_x(\partial_x A_y - \partial_y A_x)]\hat{\boldsymbol{y}} \\ &+ [E_x(\partial_z A_x - \partial_x A_z) - E_y(\partial_y A_z - \partial_z A_y)]\hat{\boldsymbol{z}}\end{aligned}$$

$$= E_x[(\partial_x A_x)\hat{\boldsymbol{x}} + (\partial_y A_x)\hat{\boldsymbol{y}} + (\partial_z A_x)\hat{\boldsymbol{z}}] + E_y[(\partial_x A_y)\hat{\boldsymbol{x}} + (\partial_y A_y)\hat{\boldsymbol{y}} + (\partial_z A_y)\hat{\boldsymbol{z}}] + E_z[(\partial_x A_z)\hat{\boldsymbol{x}} + (\partial_y A_z)\hat{\boldsymbol{y}} + (\partial_z A_z)\hat{\boldsymbol{z}}]$$

$$-E_x[(\partial_x A_x)\hat{\boldsymbol{x}} + (\partial_x A_y)\hat{\boldsymbol{y}} + (\partial_x A_z)\hat{\boldsymbol{z}}] - E_y[(\partial_y A_x)\hat{\boldsymbol{x}} + (\partial_y A_y)\hat{\boldsymbol{y}} + (\partial_y A_z)\hat{\boldsymbol{z}}] - E_z[(\partial_z A_x)\hat{\boldsymbol{x}} + (\partial_z A_y)\hat{\boldsymbol{y}} + (\partial_z A_z)\hat{\boldsymbol{z}}]$$

$$= (E_x \boldsymbol{\nabla} A_x + E_y \boldsymbol{\nabla} A_y + E_z \boldsymbol{\nabla} A_z) - (\boldsymbol{E} \cdot \boldsymbol{\nabla})\boldsymbol{A}. \qquad (31)$$

A further rearrangement involving the contribution to the angular momentum density by the second term on the right-hand side of Eq.(31) yields

$$\begin{aligned}\boldsymbol{r} \times (\boldsymbol{E} \cdot \boldsymbol{\nabla})\boldsymbol{A} &= \boldsymbol{r} \times (E_x \partial_x \boldsymbol{A} + E_y \partial_y \boldsymbol{A} + E_z \partial_z \boldsymbol{A}) \\ &= E_x[\partial_x(\boldsymbol{r} \times \boldsymbol{A}) - \hat{\boldsymbol{x}} \times \boldsymbol{A}] + E_y[\partial_y(\boldsymbol{r} \times \boldsymbol{A}) - \hat{\boldsymbol{y}} \times \boldsymbol{A}] + E_z[\partial_z(\boldsymbol{r} \times \boldsymbol{A}) - \hat{\boldsymbol{z}} \times \boldsymbol{A}] \\ &= (\boldsymbol{E} \cdot \boldsymbol{\nabla})(\boldsymbol{r} \times \boldsymbol{A}) - \boldsymbol{E} \times \boldsymbol{A}. \qquad (32)\end{aligned}$$

Substituting Eqs.(31) and (32) into the integrand of Eq.(30), we arrive at

$$\boldsymbol{\mathcal{J}}(t) = \varepsilon_0 \iiint_{-\infty}^{\infty} [(\textstyle\sum_{a=x,y,z} E_a \boldsymbol{r} \times \boldsymbol{\nabla} A_a) - (\boldsymbol{E} \cdot \boldsymbol{\nabla})(\boldsymbol{r} \times \boldsymbol{A}) + \boldsymbol{E} \times \boldsymbol{A}]\mathrm{d}v. \qquad (33)$$

The middle term of the above integral, evaluated via integration by parts, becomes

$$\iiint_{-\infty}^{\infty} (\boldsymbol{E} \cdot \boldsymbol{\nabla})(\boldsymbol{r} \times \boldsymbol{A})\mathrm{d}v = -\iiint_{-\infty}^{\infty} (\partial_x E_x + \partial_y E_y + \partial_z E_z)\,(\boldsymbol{r} \times \boldsymbol{A})\mathrm{d}v = -\iiint_{-\infty}^{\infty} (\boldsymbol{\nabla} \cdot \boldsymbol{E})(\boldsymbol{r} \times \boldsymbol{A})\mathrm{d}v. \quad (34)$$

If we now confine our attention to the transverse component $\boldsymbol{E}_\perp$ of the $E$-field by setting $\boldsymbol{\nabla}\cdot\boldsymbol{E}=0$, we can drop the middle term from the integrand of Eq.(33). (It will be shown further below that this middle term is solely responsible for the contribution of the longitudinal $E$-field $\boldsymbol{E}_\parallel$ to the overall angular momentum.) As for the vector potential $\boldsymbol{A}$, it should be clear that only the transverse component $\boldsymbol{A}_\perp$ has participated in the preceding equations, considering that we started by setting $\boldsymbol{B}=\boldsymbol{\nabla}\times\boldsymbol{A}$, which is the same as $\boldsymbol{B}=\boldsymbol{\nabla}\times\boldsymbol{A}_\perp$. Consequently, the contribution of the transverse component of the fields to the EM angular momentum may be written as

$$\boldsymbol{\mathcal{J}}(t)=\varepsilon_0\iiint_{-\infty}^{\infty}\left[\left(\textstyle\sum_{a=x,y,z}E_{\perp a}\boldsymbol{r}\times\boldsymbol{\nabla}A_{\perp a}\right)+\boldsymbol{E}_\perp\times\boldsymbol{A}_\perp\right]\mathrm{d}v. \tag{35}$$

The two terms appearing in the above expression represent, respectively, the orbital and spin angular momenta of the transverse fields of the wavepacket. Upon Fourier transforming the spatial profiles of the $\boldsymbol{E}$ and $\boldsymbol{A}$ fields using standard procedure, namely,

$$\widetilde{\boldsymbol{E}}(\boldsymbol{k},t)=\iiint_{-\infty}^{\infty}\boldsymbol{E}(\boldsymbol{r},t)e^{-\mathrm{i}\boldsymbol{k}\cdot\boldsymbol{r}}\mathrm{d}v \quad\rightarrow\quad \boldsymbol{E}(\boldsymbol{r},t)=(2\pi)^{-3}\iiint_{-\infty}^{\infty}\widetilde{\boldsymbol{E}}(\boldsymbol{k},t)e^{\mathrm{i}\boldsymbol{k}\cdot\boldsymbol{r}}\mathrm{d}\boldsymbol{k}, \tag{36}$$

the spin angular momentum part of Eq.(35) transforms to the $k$-space, as follows:

$$\begin{aligned}\boldsymbol{\mathcal{S}}(t)&=\varepsilon_0\iiint_{-\infty}^{\infty}\boldsymbol{E}_\perp(\boldsymbol{r},t)\times\boldsymbol{A}_\perp(\boldsymbol{r},t)\mathrm{d}v \quad \leftarrow \boxed{\mathrm{d}v \text{ stands for } \mathrm{d}x\mathrm{d}y\mathrm{d}z}\\ &=\varepsilon_0(2\pi)^{-6}\int_{-\infty}^{\infty}\widetilde{\boldsymbol{E}}_\perp(\boldsymbol{k},t)\times\widetilde{\boldsymbol{A}}_\perp(\boldsymbol{k}',t)\int_{-\infty}^{\infty}e^{\mathrm{i}(\boldsymbol{k}+\boldsymbol{k}')\cdot\boldsymbol{r}}\mathrm{d}v\mathrm{d}\boldsymbol{k}\mathrm{d}\boldsymbol{k}'\\ &=\varepsilon_0(2\pi)^{-3}\int_{-\infty}^{\infty}\widetilde{\boldsymbol{E}}_\perp(\boldsymbol{k},t)\times\widetilde{\boldsymbol{A}}_\perp(\boldsymbol{k}',t)\delta(\boldsymbol{k}+\boldsymbol{k}')\mathrm{d}\boldsymbol{k}'\mathrm{d}\boldsymbol{k}\\ &=\varepsilon_0(2\pi)^{-3}\int_{-\infty}^{\infty}\widetilde{\boldsymbol{E}}_\perp(\boldsymbol{k},t)\times\widetilde{\boldsymbol{A}}_\perp(-\boldsymbol{k},t)\mathrm{d}\boldsymbol{k} \quad \leftarrow \boxed{\text{Sifting property of } \delta(\boldsymbol{k}+\boldsymbol{k}')}\\ &=\varepsilon_0(2\pi)^{-3}\iiint_{-\infty}^{\infty}\widetilde{\boldsymbol{E}}_\perp(\boldsymbol{k},t)\times\widetilde{\boldsymbol{A}}_\perp^*(\boldsymbol{k},t)\mathrm{d}\boldsymbol{k}.\end{aligned} \tag{37}$$

---

**Digression**: A single, right-circularly-polarized (RCP) plane-wave with $\boldsymbol{E}(\boldsymbol{r},t)=E_{0\boldsymbol{k}}e^{\mathrm{i}(\boldsymbol{k}\cdot\boldsymbol{r}-\omega t)}\hat{\boldsymbol{e}}$ and $\boldsymbol{A}(\boldsymbol{r},t)=-(\mathrm{i}E_{0\boldsymbol{k}}/\omega)e^{\mathrm{i}(\boldsymbol{k}\cdot\boldsymbol{r}-\omega t)}\hat{\boldsymbol{e}}$ will have

$$\varepsilon_0\widetilde{\boldsymbol{E}}_\perp\times\widetilde{\boldsymbol{A}}_\perp^*=\mathrm{i}(\varepsilon_0V^2E_{0\boldsymbol{k}}E_{0\boldsymbol{k}}^*/\omega)(\boldsymbol{e}'+\mathrm{i}\boldsymbol{e}'')\times(\boldsymbol{e}'-\mathrm{i}\boldsymbol{e}'')=(2\varepsilon_0V^2|E_{0\boldsymbol{k}}|^2/\omega)\boldsymbol{e}'\times\boldsymbol{e}''. \tag{38}$$

Recall that $\mathrm{d}\boldsymbol{k}=(2\pi/L)^3$, where $L^3=V$ is the volume occupied by the plane-wave. One must also take into account the contribution of the complex conjugates of the $\boldsymbol{E}$ and $\boldsymbol{A}$ fields, which doubles the value of the spin angular momentum given by Eq.(38). Considering that, for an RCP plane-wave, $|\boldsymbol{e}'|=|\boldsymbol{e}''|=1/\sqrt{2}$, and that $|E_{0\boldsymbol{k}}|^2=\hbar\omega/(2\varepsilon_0V)$ for a single photon of frequency $\omega$, the RCP photon is seen to have a spin angular momentum of $\hbar$ in the $\hat{\boldsymbol{\kappa}}$ direction.

---

Expressing the cross-product of $\widetilde{\boldsymbol{E}}_\perp$ and $\widetilde{\boldsymbol{A}}_\perp^*$ in Eq.(37) in terms of the normal variables $\alpha_\ell(t)$ and $\alpha_\ell^*(t)$ requires that the polarization of each plane-wave in the $k$-space be described as a superposition of right- and left-circular polarizations $\hat{\boldsymbol{e}}_{1,2}=\boldsymbol{e}'\pm\mathrm{i}\boldsymbol{e}''$, where $|\boldsymbol{e}'|=|\boldsymbol{e}''|=1/\sqrt{2}$ and $\boldsymbol{e}'\cdot\boldsymbol{e}''=0$. In this way, we will have $\hat{\boldsymbol{e}}_1\times\hat{\boldsymbol{e}}_1^*=-\mathrm{i}\hat{\boldsymbol{\kappa}}$ and $\hat{\boldsymbol{e}}_2\times\hat{\boldsymbol{e}}_2^*=\mathrm{i}\hat{\boldsymbol{\kappa}}$, whereas $\hat{\boldsymbol{e}}_1\times\hat{\boldsymbol{e}}_2^*=\hat{\boldsymbol{e}}_2\times\hat{\boldsymbol{e}}_1^*=0$, which enables us to rewrite Eq.(37) as follows:

$$\begin{aligned}\boldsymbol{\mathcal{S}}(t)&=\varepsilon_0L^{-3}\textstyle\sum_\ell\tilde{E}_{\perp,\ell}\tilde{A}_{\perp,\ell}^*(\mp\mathrm{i}\hat{\boldsymbol{\kappa}}_\ell) \quad \leftarrow \boxed{\text{upper or lower sign, depending on whether mode } \ell \text{ is RCP or LCP}}\\ &=\varepsilon_0L^{-3}\textstyle\sum_\ell[(\omega_\ell\tilde{A}_{\perp,\ell}-\mathrm{i}\tilde{E}_{\perp,\ell})(\omega_\ell\tilde{A}_{\perp,\ell}^*+\mathrm{i}\tilde{E}_{\perp,\ell}^*)-(\omega_\ell^2\tilde{A}_{\perp,\ell}\tilde{A}_{\perp,\ell}^*+\tilde{E}_{\perp,\ell}\tilde{E}_{\perp,\ell}^*)](\pm\hat{\boldsymbol{\kappa}}_\ell/2\omega_\ell)\end{aligned}$$

$\boxed{\sum_\ell\tilde{E}_{\perp,\ell}\tilde{A}_{\perp,\ell}^*\hat{\boldsymbol{\kappa}}_\ell=\sum_\ell\tilde{E}_{\perp,-\ell}\tilde{A}_{\perp,-\ell}^*\hat{\boldsymbol{\kappa}}_{-\ell}=-\sum_\ell\tilde{E}_{\perp,\ell}^*\tilde{A}_{\perp,\ell}\hat{\boldsymbol{\kappa}}_\ell}$ $\boxed{\text{This term is the same for } \ell \text{ and } -\ell\text{, but } \hat{\boldsymbol{\kappa}}_\ell \text{ switches sign}}$

$$= \varepsilon_0 L^{-3} \textstyle\sum_\ell \alpha_\ell(t)\alpha_\ell^*(t)(\pm\hat{\boldsymbol{\kappa}}_\ell/2\zeta_\ell^2\omega_\ell)$$

$$= \textstyle\sum_\ell [\underbrace{\alpha_{\ell(+)}(t)\alpha_{\ell(+)}^*(t)}_{\text{mode } \ell \text{ is RCP}} - \underbrace{\alpha_{\ell(-)}(t)\alpha_{\ell(-)}^*(t)}_{\text{mode } \ell \text{ is LCP}}]\hbar\hat{\boldsymbol{\kappa}}_\ell. \qquad (39)$$

It is thus seen to be possible to define creation and annihilation operators $\hat{a}$ and $\hat{a}^\dagger$ (corresponding to $\alpha$ and $\alpha^*$) for the spin angular momentum of EM wavepackets provided that the orthogonal polarization states for each mode are taken to be the right- and left-circular states.

The situation is much more complicated for the EM orbital angular momentum, where, upon transformation to the Fourier domain, the first term of Eq.(35) becomes

$$\boldsymbol{\mathcal{L}}(t) = \varepsilon_0 \iiint_{-\infty}^{\infty} [\textstyle\sum_{a=x,y,z} E_{\perp a}(\boldsymbol{r},t)\boldsymbol{r} \times \boldsymbol{\nabla} A_{\perp a}(\boldsymbol{r},t)]\mathrm{d}v \leftarrow \boxed{\mathrm{d}v \text{ stands for } \mathrm{d}x\mathrm{d}y\mathrm{d}z}$$

$$= \varepsilon_0(2\pi)^{-6}\textstyle\sum_{a=x,y,z}\int_{-\infty}^{\infty} \tilde{E}_{\perp a}(\boldsymbol{k}',t)e^{\mathrm{i}\boldsymbol{k}'\cdot\boldsymbol{r}}\boldsymbol{r} \times \mathrm{i}\boldsymbol{k}\tilde{A}_{\perp a}(\boldsymbol{k},t)e^{\mathrm{i}\boldsymbol{k}\cdot\boldsymbol{r}}\mathrm{d}\boldsymbol{k}'\mathrm{d}\boldsymbol{k}\mathrm{d}v$$

$$= \varepsilon_0(2\pi)^{-6}\textstyle\sum_{a=x,y,z}\int_{-\infty}^{\infty} \tilde{E}_{\perp a}(\boldsymbol{k}',t)[\int_{-\infty}^{\infty}\mathrm{i}\boldsymbol{r}e^{\mathrm{i}(\boldsymbol{k}+\boldsymbol{k}')\cdot\boldsymbol{r}}\mathrm{d}v] \times \boldsymbol{k}\tilde{A}_{\perp a}(\boldsymbol{k},t)\mathrm{d}\boldsymbol{k}\mathrm{d}\boldsymbol{k}'$$

$$= \varepsilon_0(2\pi)^{-3}\textstyle\sum_{a=x,y,z}\int_{-\infty}^{\infty} \tilde{E}_{\perp a}(\boldsymbol{k}',t)[\boldsymbol{\nabla}_{\boldsymbol{k}}\delta(\boldsymbol{k}+\boldsymbol{k}')] \times \boldsymbol{k}\tilde{A}_{\perp a}(\boldsymbol{k},t)\mathrm{d}\boldsymbol{k}\mathrm{d}\boldsymbol{k}'$$

$$= \varepsilon_0(2\pi)^{-3}\textstyle\sum_{a=x,y,z}\int_{-\infty}^{\infty} \tilde{E}_{\perp a}(\boldsymbol{k}',t)\{\boldsymbol{\nabla}_{\boldsymbol{k}} \times \underbrace{[\delta(\boldsymbol{k}+\boldsymbol{k}')\boldsymbol{k}\tilde{A}_{\perp a}(\boldsymbol{k},t)]}_{0} \leftarrow \boxed{\text{Partial derivatives with respect to } k_x, k_y, k_z \text{ integrate to 0 over } k\text{-space}}$$

$$\boxed{\boldsymbol{\nabla} \times (\psi\boldsymbol{A}) = (\boldsymbol{\nabla}\psi) \times \boldsymbol{A} + \psi\boldsymbol{\nabla} \times \boldsymbol{A}} \rightarrow \quad -\delta(\boldsymbol{k}+\boldsymbol{k}')\boldsymbol{\nabla}_{\boldsymbol{k}} \times [\boldsymbol{k}\tilde{A}_{\perp a}(\boldsymbol{k},t)]\}\mathrm{d}\boldsymbol{k}\mathrm{d}\boldsymbol{k}'$$

$$= \varepsilon_0(2\pi)^{-3}\textstyle\sum_{a=x,y,z}\int_{-\infty}^{\infty} \tilde{E}_{\perp a}(\boldsymbol{k}',t)\delta(\boldsymbol{k}+\boldsymbol{k}')[\boldsymbol{k} \times \boldsymbol{\nabla}_{\boldsymbol{k}}\tilde{A}_{\perp a}(\boldsymbol{k},t) - \tilde{A}_{\perp a}(\boldsymbol{k},t)\underbrace{\boldsymbol{\nabla}_{\boldsymbol{k}} \times \boldsymbol{k}}_{0}]\mathrm{d}\boldsymbol{k}'\mathrm{d}\boldsymbol{k}$$

$$= \varepsilon_0(2\pi)^{-3}\textstyle\sum_{a=x,y,z}\int_{-\infty}^{\infty} \tilde{E}_{\perp a}(-\boldsymbol{k},t)\boldsymbol{k} \times \boldsymbol{\nabla}_{\boldsymbol{k}}\tilde{A}_{\perp a}(\boldsymbol{k},t)\mathrm{d}\boldsymbol{k} \leftarrow \boxed{\text{Sifting property of } \delta(\boldsymbol{k}+\boldsymbol{k}')}$$

$$= \varepsilon_0(2\pi)^{-3}\textstyle\sum_{a=x,y,z}\int_{-\infty}^{\infty} \tilde{E}_{\perp a}^*(\boldsymbol{k},t)\boldsymbol{k} \times \boldsymbol{\nabla}_{\boldsymbol{k}}\tilde{A}_{\perp a}(\boldsymbol{k},t)\mathrm{d}\boldsymbol{k}. \leftarrow \boxed{\mathrm{d}\boldsymbol{k} \text{ stands for } \mathrm{d}k_x\mathrm{d}k_y\mathrm{d}k_z} \qquad (40)$$

We have thus found a compact expression for the orbital angular momentum of the transverse fields of an EM wavepacket.[1] While this $k$-space integral is intimately tied to the propagating plane-wave modes that constitute the packet, it is not immediately clear how it could be written in terms of the normal variables $\alpha_\ell(t)$ and $\alpha_\ell^*(t)$, which is perhaps a necessary step prior to quantization. The interested reader may want to consult Complement $B_I$ in Ref.[1] for additional insights.

---

**Digression**: Returning to Eq.(33) and seeking to evaluate the angular momentum associated with the longitudinal component of the $E$-field, we find that the 1st term of the integral is cancelled by its 3rd term. This is readily confirmed in the Coulomb gauge, where $\boldsymbol{A}_\parallel(\boldsymbol{r},t) = 0$ and $\boldsymbol{E}_\parallel(\boldsymbol{r},t) = -\boldsymbol{\nabla}\psi(\boldsymbol{r},t)$, and, therefore,

$$\int_{-\infty}^{\infty}[(\textstyle\sum_{a=x,y,z} E_{\parallel a}\boldsymbol{r} \times \boldsymbol{\nabla}A_{\perp a}) + \boldsymbol{E}_\parallel \times \boldsymbol{A}_\perp]\mathrm{d}v = -\int_{-\infty}^{\infty}\{[\textstyle\sum_{a=x,y,z}(\partial_a\psi)\boldsymbol{r} \times \boldsymbol{\nabla}A_{\perp a}] + \boldsymbol{\nabla}\psi \times \boldsymbol{A}_\perp\}\mathrm{d}v$$

$$\boxed{\begin{array}{c}\text{use integration by parts;}\\ \text{also } \boldsymbol{\nabla} \times (\psi\boldsymbol{A}) = (\boldsymbol{\nabla}\psi) \times \boldsymbol{A} + \psi\boldsymbol{\nabla} \times \boldsymbol{A}\end{array}} \rightarrow = \int_{-\infty}^{\infty}\{[\psi\textstyle\sum_{a=x,y,z}\partial_a(\boldsymbol{r} \times \boldsymbol{\nabla}A_{\perp a})] + \psi\boldsymbol{\nabla} \times \boldsymbol{A}_\perp\}\mathrm{d}v. \qquad (41)$$

Simplifying the first term of the integrand appearing on the right-hand side of Eq.(41), we find

$$\textstyle\sum_{a=x,y,z}\partial_a(\boldsymbol{r} \times \boldsymbol{\nabla}A_{\perp a})$$

$$= (\hat{\boldsymbol{x}} \times \boldsymbol{\nabla}A_{\perp x} + \boldsymbol{r} \times \partial_x\boldsymbol{\nabla}A_{\perp x}) + (\hat{\boldsymbol{y}} \times \boldsymbol{\nabla}A_{\perp y} + \boldsymbol{r} \times \partial_y\boldsymbol{\nabla}A_{\perp y}) + (\hat{\boldsymbol{z}} \times \boldsymbol{\nabla}A_{\perp z} + \boldsymbol{r} \times \partial_z\boldsymbol{\nabla}A_{\perp z})$$

$$= (\partial_y A_{\perp x}\hat{\boldsymbol{z}} - \partial_z A_{\perp x}\hat{\boldsymbol{y}}) + (\partial_z A_{\perp y}\hat{\boldsymbol{x}} - \partial_x A_{\perp y}\hat{\boldsymbol{z}}) + (\partial_x A_{\perp z}\hat{\boldsymbol{y}} - \partial_y A_{\perp z}\hat{\boldsymbol{x}})$$
$$+\boldsymbol{r} \times \partial_x(\partial_x A_{\perp x}\hat{\boldsymbol{x}} + \partial_y A_{\perp x}\hat{\boldsymbol{y}} + \partial_z A_{\perp x}\hat{\boldsymbol{z}}) + \boldsymbol{r} \times \partial_y(\partial_x A_{\perp y}\hat{\boldsymbol{x}} + \partial_y A_{\perp y}\hat{\boldsymbol{y}} + \partial_z A_{\perp y}\hat{\boldsymbol{z}})$$
$$+\boldsymbol{r} \times \partial_z(\partial_x A_{\perp z}\hat{\boldsymbol{x}} + \partial_y A_{\perp z}\hat{\boldsymbol{y}} + \partial_z A_{\perp z}\hat{\boldsymbol{z}}) = -\boldsymbol{\nabla} \times \boldsymbol{A}_\perp + \boldsymbol{r} \times \boldsymbol{\nabla}(\partial_x A_{\perp x} + \partial_y A_{\perp y} + \partial_z A_{\perp z})$$
$$= -\boldsymbol{\nabla} \times \boldsymbol{A}_\perp + \boldsymbol{r} \times \boldsymbol{\nabla}(\cancelto{0}{\boldsymbol{\nabla} \cdot \boldsymbol{A}_\perp}) = -\boldsymbol{\nabla} \times \boldsymbol{A}_\perp. \tag{42}$$

Thus, the integrand of Eq.(41) vanishes. Substituting from Eq.(34) into the remaining part of Eq.(33) and using the fact that $\varepsilon_0 \boldsymbol{\nabla} \cdot \boldsymbol{E}_\parallel = \rho(\boldsymbol{r}, t) = \sum_n q_n \delta[\boldsymbol{r} - \boldsymbol{r}_n(t)]$, we finally arrive at

$$\boldsymbol{\mathcal{J}}(t) = \varepsilon_0 \iiint_{-\infty}^{\infty} (\boldsymbol{\nabla} \cdot \boldsymbol{E}_\parallel)(\boldsymbol{r} \times \boldsymbol{A}_\perp)\mathrm{d}v = \iiint_{-\infty}^{\infty} \sum_n q_n \delta[\boldsymbol{r} - \boldsymbol{r}_n(t)]\,(\boldsymbol{r} \times \boldsymbol{A}_\perp)\mathrm{d}v$$
$$= \sum_n q_n \boldsymbol{r}_n(t) \times \boldsymbol{A}_\perp[\boldsymbol{r}_n(t), t]. \tag{43}$$

This, of course, is the same result as the one found in Sec.4, Eq.(12), using an alternative approach.

**10. Concluding remarks**. This paper has brought together the various elements of the classical theory of electrodynamics as needed for an understanding of the quantization of the electromagnetic field in the context of the quantum theory of radiation. Starting with a collection of electric point-charges moving arbitrarily in free space, we showed that the longitudinal component $\boldsymbol{E}_\parallel(\boldsymbol{r}, t)$ of the electric field can be relegated to certain characteristics of the EM potentials that are localized on the point-charges and can, therefore, be split off from the radiation field and bundled together with the dynamic properties (i.e., kinetic energies and mechanical momenta) of the point-charges. The transverse components $\boldsymbol{E}_\perp(\boldsymbol{r}, t)$ and $\boldsymbol{B}(\boldsymbol{r}, t)$ of the radiation field were subsequently discretized and associated with propagating plane-waves (or modes), specified in terms of their frequency $\omega$, propagation vector $\boldsymbol{k}$, polarization vector $\hat{\boldsymbol{e}}$, and a so-called normal variable $\alpha(t)$ as well as its complex conjugate $\alpha^*(t)$. The normal variable $\alpha(t)$ is the precursor of the annihilation operator $\hat{a}$ assigned to the $(\omega, \boldsymbol{k}, \hat{\boldsymbol{e}})$ mode, while $\alpha^*(t)$ presages the corresponding creation operator $\hat{a}^\dagger$.

In the second half of the paper, we examined the energy and the linear momentum, and also the spin and orbital angular momenta of the radiation field, keeping the multi-modal structure of the field front and center. We found that the first three characteristics (i.e., energy content, linear momentum, and spin angular momentum) are neatly divisible among the various modes in proportion to the corresponding product $\alpha(t)\alpha^*(t)$ for each mode — independently of all the other modes. The orbital angular momentum, however, while expressible as the integral of a compact expression over all propagating plane-wave modes (shown in Eq.(40)), could not be divided among the individual modes, requiring instead that the multimodal structure of the host wavepacket in its entirety be taken into account.

In the case of free EM wavepackets, where $\boldsymbol{J}_{\text{free}}(\boldsymbol{r}, t) = 0$, Eq.(14) and its descendants become simpler, leading to $\alpha_\ell(t) = e^{-\mathrm{i}\omega_\ell t}$ and $\tilde{E}_{\perp,\ell}(t) = \mathrm{i}\omega_\ell \tilde{A}_{\perp,\ell}(t) = \mathrm{i}L^3\sqrt{\hbar\omega_\ell/(2\varepsilon_0 V)}\,e^{-\mathrm{i}\omega_\ell t}$. The total EM angular momentum $\boldsymbol{\mathcal{J}} = \boldsymbol{\mathcal{L}} + \boldsymbol{\mathcal{S}}$ in this case can be computed using an alternative method,[6] but the end results turn out to be the same as those given by Eqs.(37) and (40), as shown in the Appendix.

## Appendix

### Equivalence of two expressions for the angular momentum of free EM wavepackets

In our *Phys. Rev. A* (2011) paper[6], Fourier transforming the $\boldsymbol{E}$ and $\boldsymbol{H}$ fields over $(x, y, t)$ takes these fields to the $(k_x, k_y, \omega)$ domain, where the transformed $E$-field, namely, $\boldsymbol{\mathcal{E}}(k_x, k_y, \omega)$, satisfies the following identity:

$$\boldsymbol{E}(\boldsymbol{r},t) = \int_{\omega=-\infty}^{\infty} \iint_{k_x^2+k_y^2<(\omega/c)^2} \boldsymbol{\mathcal{E}}(k_x,k_y,\omega) e^{\mathrm{i}(\boldsymbol{k}\cdot\boldsymbol{r}-\omega t)} \mathrm{d}k_x \mathrm{d}k_y \mathrm{d}\omega. \qquad \text{(A1)}$$

In contrast, when the Fourier domain variables are taken to be $(k_x,k_y,k_z)$, as in Complement $\mathrm{B_I}$ of Ref.[1], we find

$$\boldsymbol{E}(\boldsymbol{r},t) = \iiint_{-\infty}^{\infty} \widetilde{\boldsymbol{E}}(k_x,k_y,k_z) e^{\mathrm{i}(\boldsymbol{k}\cdot\boldsymbol{r}-\omega t)} \mathrm{d}k_x \mathrm{d}k_y \mathrm{d}k_z. \qquad \text{(A2)}$$

Considering that $\omega = ck = c\sqrt{k_x^2+k_y^2+k_z^2}$, if we fix $(k_x,k_y)$ and allow the change in $\omega$ to be driven solely by the change in $k_z$, we will have $\partial\omega/\partial k_z = ck_z/k$. Consequently, the differential volume element in the $(k_x,k_y,\omega)$ space is related to the corresponding volume element in the $(k_x,k_y,k_z)$ space via $\mathrm{d}k_x\mathrm{d}k_y\mathrm{d}\omega = (ck_z/k)\mathrm{d}k_x\mathrm{d}k_y\mathrm{d}k_z$. The same result, of course, may be obtained by computing the Jacobian of the transformation between the two sets of coordinates; that is,

$$\frac{\partial(k_x,k_y,\omega)}{\partial(k_x,k_y,k_z)} = \left\|\begin{matrix} \frac{\partial k_x}{\partial k_x} & \frac{\partial k_y}{\partial k_x} & \frac{\partial\omega}{\partial k_x} \\ \frac{\partial k_x}{\partial k_y} & \frac{\partial k_y}{\partial k_y} & \frac{\partial\omega}{\partial k_y} \\ \frac{\partial k_x}{\partial k_z} & \frac{\partial k_y}{\partial k_z} & \frac{\partial\omega}{\partial k_z} \end{matrix}\right\| = \left\|\begin{matrix} 1 & 0 & ck_x/k \\ 0 & 1 & ck_y/k \\ 0 & 0 & ck_z/k \end{matrix}\right\| = ck_z/k. \qquad \text{(A3)}$$

Comparing Eqs.(A1) and (A2), we conclude that $\widetilde{\boldsymbol{E}}(\boldsymbol{k}) = \widetilde{\boldsymbol{E}}(k_x,k_y,k_z) = (ck_z/k)\boldsymbol{\mathcal{E}}(k_x,k_y,\omega)$. Next, intending to relate the expression of the spin angular momentum in the $(k_x,k_y,k_z)$ domain—as given by Eq.(11) in Complement $\mathrm{B_I}$ of Ref.[1]—to that in the $(k_x,k_y,\omega)$ domain, we invoke the identity $\widetilde{\boldsymbol{E}}\cdot\boldsymbol{k}=0$ to substitute for $\tilde{E}_z$ in terms of $\tilde{E}_x$, $\tilde{E}_y$, and the three components of $\boldsymbol{k}$, as follows:

$$\begin{aligned}
\widetilde{\boldsymbol{E}}(\boldsymbol{k}) \times \widetilde{\boldsymbol{E}}^*(\boldsymbol{k}) &= (\tilde{E}_x\hat{\boldsymbol{x}} + \tilde{E}_y\hat{\boldsymbol{y}} + \tilde{E}_z\hat{\boldsymbol{z}}) \times (\tilde{E}_x^*\hat{\boldsymbol{x}} + \tilde{E}_y^*\hat{\boldsymbol{y}} + \tilde{E}_z^*\hat{\boldsymbol{z}}) \\
&= \{\tilde{E}_x\hat{\boldsymbol{x}} + \tilde{E}_y\hat{\boldsymbol{y}} - [(k_x/k_z)\tilde{E}_x + (k_y/k_z)\tilde{E}_y]\hat{\boldsymbol{z}}\} \times \{\tilde{E}_x^*\hat{\boldsymbol{x}} + \tilde{E}_y^*\hat{\boldsymbol{y}} - [(k_x/k_z)\tilde{E}_x^* + (k_y/k_z)\tilde{E}_y^*]\hat{\boldsymbol{z}}\} \\
&= \tilde{E}_x\tilde{E}_y^*\hat{\boldsymbol{z}} + [(k_x/k_z)\tilde{E}_x\tilde{E}_x^* + (k_y/k_z)\tilde{E}_x\tilde{E}_y^*]\hat{\boldsymbol{y}} - \tilde{E}_y\tilde{E}_x^*\hat{\boldsymbol{z}} - [(k_x/k_z)\tilde{E}_y\tilde{E}_x^* + (k_y/k_z)\tilde{E}_y\tilde{E}_y^*]\hat{\boldsymbol{x}} \\
&\quad -[(k_x/k_z)\tilde{E}_x\tilde{E}_x^* + (k_y/k_z)\tilde{E}_y\tilde{E}_x^*]\hat{\boldsymbol{y}} + [(k_x/k_z)\tilde{E}_x\tilde{E}_y^* + (k_y/k_z)\tilde{E}_y\tilde{E}_y^*]\hat{\boldsymbol{x}} \\
&= (k_x/k_z)(\tilde{E}_x\tilde{E}_y^* - \tilde{E}_y\tilde{E}_x^*)\hat{\boldsymbol{x}} + (k_y/k_z)(\tilde{E}_x\tilde{E}_y^* - \tilde{E}_y\tilde{E}_x^*)\hat{\boldsymbol{y}} + (\tilde{E}_x\tilde{E}_y^* - \tilde{E}_y\tilde{E}_x^*)\hat{\boldsymbol{z}} \\
&= (\tilde{E}_x\tilde{E}_y^* - \tilde{E}_y\tilde{E}_x^*)(k_x\hat{\boldsymbol{x}} + k_y\hat{\boldsymbol{y}} + k_z\hat{\boldsymbol{z}})/k_z = (k/k_z)(\tilde{E}_x\tilde{E}_y^* - \tilde{E}_y\tilde{E}_x^*)\hat{\boldsymbol{\kappa}}. \qquad \text{(A4)}
\end{aligned}$$

The spin angular momentum may now be written as

$$\begin{aligned}
\boldsymbol{\mathcal{S}} &= \mathrm{i}\varepsilon_0\{\iiint_{-\infty}^{\infty} \omega^{-1}\widetilde{\boldsymbol{E}}(\boldsymbol{k}) \times \widetilde{\boldsymbol{E}}^*(\boldsymbol{k})\mathrm{d}k_x\mathrm{d}k_y\mathrm{d}k_z\} = \mathrm{i}\varepsilon_0\{\iiint_{-\infty}^{\infty} \omega^{-1}(\tilde{E}_x\tilde{E}_y^* - \tilde{E}_y\tilde{E}_x^*)(\boldsymbol{k}/k_z)\mathrm{d}k_x\mathrm{d}k_y\mathrm{d}k_z\} \\
&= \mathrm{i}\varepsilon_0\{\iiint_{-\infty}^{\infty} \omega^{-1}(ck_z/k)^2(\mathcal{E}_x\mathcal{E}_y^* - \mathcal{E}_y\mathcal{E}_x^*)(\boldsymbol{k}/k_z)(ck_z/k)^{-1}\mathrm{d}k_x\mathrm{d}k_y\mathrm{d}\omega\} \\
&= \mathrm{i}\varepsilon_0\{\iiint_{-\infty}^{\infty} (c/\omega)^2(\mathcal{E}_x\mathcal{E}_y^* - \mathcal{E}_y\mathcal{E}_x^*)\boldsymbol{k}\,\mathrm{d}k_x\mathrm{d}k_y\mathrm{d}\omega\}. \qquad \text{(A5)}
\end{aligned}$$

The last expression is the same as that appearing within Eq.(12) of our *Phys. Rev. A* (2011) paper.[6]

With regard to the orbital angular momentum $\boldsymbol{\mathcal{L}}$, we note that $\omega = ck = c(k_x^2+k_y^2+k_z^2)^{½}$ changes by $\Delta\omega = (\partial\omega/\partial k_x)\Delta k_x + (\partial\omega/\partial k_z)\Delta k_z = c(k_x\Delta k_x + k_z\Delta k_z)/k$ if $k_x$ changes by $\Delta k_x$ and $k_z$ by $\Delta k_z$, while $k_y$ is kept constant. Consequently, $\Delta k_y = 0$ and $\Delta\omega = 0$ together imply that $\Delta k_z = -(k_x/k_z)\Delta k_x$. Thus, upon varying $k_x$ and $k_z$ while fixing $k_y$ and $\omega$, we arrive at

$$\Delta\tilde{\boldsymbol{E}}(k_x,k_y,k_z) = (\partial_{k_x}\tilde{\boldsymbol{E}})\Delta k_x + (\partial_{k_z}\tilde{\boldsymbol{E}})\Delta k_z = [\partial_{k_x}\tilde{\boldsymbol{E}}(\boldsymbol{k}) - (k_x/k_z)\partial_{k_z}\tilde{\boldsymbol{E}}(\boldsymbol{k})]\Delta k_x$$

$$\rightarrow \quad \partial_{k_x}\tilde{\boldsymbol{E}}(k_x,k_y,\omega) = \partial_{k_x}\tilde{\boldsymbol{E}}(\boldsymbol{k}) - (k_x/k_z)\partial_{k_z}\tilde{\boldsymbol{E}}(\boldsymbol{k}). \tag{A6}$$

Similarly, varying $k_y$ and $k_z$ (while keeping $k_x$ and $\omega$ fixed) yields

$$\partial_{k_y}\tilde{\boldsymbol{E}}(k_x,k_y,\omega) = \partial_{k_y}\tilde{\boldsymbol{E}}(\boldsymbol{k}) - (k_y/k_z)\partial_{k_z}\tilde{\boldsymbol{E}}(\boldsymbol{k}). \tag{A7}$$

Now, the term pertaining to the orbital angular momentum $\boldsymbol{\mathcal{L}}$ appearing in the expression of total angular momentum $\boldsymbol{\mathcal{J}}$ within Eq.(12) of our *Phys. Rev. A* (2011) paper[6] may be manipulated as follows:

$$\begin{aligned}
&\{[\tilde{\boldsymbol{E}}\cdot\partial_{k_x}\tilde{\boldsymbol{E}}^*(k_x,k_y,\omega)]\hat{\boldsymbol{x}} + [\tilde{\boldsymbol{E}}\cdot\partial_{k_y}\tilde{\boldsymbol{E}}^*(k_x,k_y,\omega)]\hat{\boldsymbol{y}}\}\times\boldsymbol{k}\\
&= \{[\tilde{\boldsymbol{E}}\cdot\partial_{k_x}\tilde{\boldsymbol{E}}^*(\boldsymbol{k}) - (k_x/k_z)\tilde{\boldsymbol{E}}\cdot\partial_{k_z}\tilde{\boldsymbol{E}}^*(\boldsymbol{k})]\hat{\boldsymbol{x}} + [\tilde{\boldsymbol{E}}\cdot\partial_{k_y}\tilde{\boldsymbol{E}}^*(\boldsymbol{k}) - (k_y/k_z)\tilde{\boldsymbol{E}}\cdot\partial_{k_z}\tilde{\boldsymbol{E}}^*(\boldsymbol{k})]\hat{\boldsymbol{y}}\}\times\boldsymbol{k}\\
&= \{[\tilde{\boldsymbol{E}}\cdot\partial_{k_x}\tilde{\boldsymbol{E}}^*(\boldsymbol{k})]\hat{\boldsymbol{x}} + [\tilde{\boldsymbol{E}}\cdot\partial_{k_y}\tilde{\boldsymbol{E}}^*(\boldsymbol{k})]\hat{\boldsymbol{y}} - [\tilde{\boldsymbol{E}}\cdot\partial_{k_z}\tilde{\boldsymbol{E}}^*(\boldsymbol{k})]\,(k_x\hat{\boldsymbol{x}} + k_y\hat{\boldsymbol{y}} + k_z\hat{\boldsymbol{z}} - k_z\hat{\boldsymbol{z}})/k_z\}\times\boldsymbol{k} \quad \boxed{\boldsymbol{k}\times\boldsymbol{k}=0}\\
&= \{[\tilde{\boldsymbol{E}}\cdot\partial_{k_x}\tilde{\boldsymbol{E}}^*(\boldsymbol{k})]\hat{\boldsymbol{x}} + [\tilde{\boldsymbol{E}}\cdot\partial_{k_y}\tilde{\boldsymbol{E}}^*(\boldsymbol{k})]\hat{\boldsymbol{y}} + [\tilde{\boldsymbol{E}}\cdot\partial_{k_z}\tilde{\boldsymbol{E}}^*(\boldsymbol{k})]\hat{\boldsymbol{z}}\}\times\boldsymbol{k}\\
&= (\tilde{E}_x\partial_{k_x}\tilde{E}_x^* + \tilde{E}_y\partial_{k_x}\tilde{E}_y^* + \tilde{E}_z\partial_{k_x}\tilde{E}_z^*)(k_y\hat{\boldsymbol{z}} - k_z\hat{\boldsymbol{y}}) + (\tilde{E}_x\partial_{k_y}\tilde{E}_x^* + \tilde{E}_y\partial_{k_y}\tilde{E}_y^* + \tilde{E}_z\partial_{k_y}\tilde{E}_z^*)(k_z\hat{\boldsymbol{x}} - k_x\hat{\boldsymbol{z}})\\
&\qquad + (\tilde{E}_x\partial_{k_z}\tilde{E}_x^* + \tilde{E}_y\partial_{k_z}\tilde{E}_y^* + \tilde{E}_z\partial_{k_z}\tilde{E}_z^*)(k_x\hat{\boldsymbol{y}} - k_y\hat{\boldsymbol{x}})\\
&= -\tilde{E}_x(\boldsymbol{k}\times\boldsymbol{\nabla}_k\tilde{E}_x^*) - \tilde{E}_y(\boldsymbol{k}\times\boldsymbol{\nabla}_k\tilde{E}_y^*) - \tilde{E}_z(\boldsymbol{k}\times\boldsymbol{\nabla}_k\tilde{E}_z^*).
\end{aligned} \tag{A8}$$

Substitution into the standard expression for orbital angular momentum in free space (as given by the first part of Eq.(11) in Complement $\mathrm{B_I}$ of Ref.[1], with $\tilde{\boldsymbol{A}}_\perp$ equated with $-\mathrm{i}\omega^{-1}\tilde{\boldsymbol{E}}$)[‡‡] yields

$$\begin{aligned}
\boldsymbol{\mathcal{L}} &= \mathrm{i}\varepsilon_0\iiint_{-\infty}^{\infty}\omega^{-1}\big[\tilde{E}_x(\boldsymbol{k}\times\boldsymbol{\nabla}_k\tilde{E}_x^*) + \tilde{E}_y(\boldsymbol{k}\times\boldsymbol{\nabla}_k\tilde{E}_y^*) + \tilde{E}_z(\boldsymbol{k}\times\boldsymbol{\nabla}_k\tilde{E}_z^*)\big]\mathrm{d}k_x\mathrm{d}k_y\mathrm{d}k_z\\
&= -\mathrm{i}\varepsilon_0\iiint_{-\infty}^{\infty}\omega^{-1}\{[\tilde{\boldsymbol{E}}\cdot\partial_{k_x}\tilde{\boldsymbol{E}}^*(k_x,k_y,\omega)]\hat{\boldsymbol{x}} + [\tilde{\boldsymbol{E}}\cdot\partial_{k_y}\tilde{\boldsymbol{E}}^*(k_x,k_y,\omega)]\hat{\boldsymbol{y}}\}\times\boldsymbol{k}\,\mathrm{d}k_x\mathrm{d}k_y\mathrm{d}k_z.
\end{aligned} \tag{A9}$$

Considering that $\tilde{\boldsymbol{E}}(k_x,k_y,\omega) = (ck_z/k)\boldsymbol{\mathcal{E}}(k_x,k_y,\omega)$, and that $\partial k_z/\partial k_x = -(k_x/k_z)$, we find

$$\begin{aligned}
\tilde{\boldsymbol{E}}\cdot\partial_{k_x}\tilde{\boldsymbol{E}}^* &= (ck_z/k)\boldsymbol{\mathcal{E}}\cdot\partial_{k_x}[(ck_z/k)\boldsymbol{\mathcal{E}}^*] = (ck_z/k)(c/k)(-k_x/k_z)\boldsymbol{\mathcal{E}}\cdot\boldsymbol{\mathcal{E}}^* + (ck_z/k)^2\boldsymbol{\mathcal{E}}\cdot\partial_{k_x}\boldsymbol{\mathcal{E}}^*\\
&= (ck_z/k)^2\boldsymbol{\mathcal{E}}\cdot\partial_{k_x}\boldsymbol{\mathcal{E}}^* - (c/k)^2k_x\boldsymbol{\mathcal{E}}\cdot\boldsymbol{\mathcal{E}}^*.
\end{aligned} \tag{A10}$$

Similarly, given that $\partial k_z/\partial k_y = -(k_y/k_z)$, we may write

$$\tilde{\boldsymbol{E}}\cdot\partial_{k_y}\tilde{\boldsymbol{E}}^*(k_x,k_y,\omega) = (ck_z/k)\boldsymbol{\mathcal{E}}\cdot\partial_{k_y}[(ck_z/k)\boldsymbol{\mathcal{E}}^*] = (ck_z/k)^2\boldsymbol{\mathcal{E}}\cdot\partial_{k_y}\boldsymbol{\mathcal{E}}^* - (c/k)^2k_y\boldsymbol{\mathcal{E}}\cdot\boldsymbol{\mathcal{E}}^*. \tag{A11}$$

Upon substituting from Eqs.(A10) and (A11) into Eq.(A9), we finally arrive at

---

[‡‡] For electromagnetic plane-waves in free space, where $\tilde{\boldsymbol{E}}(\boldsymbol{k}) = \mathrm{i}\omega\tilde{\boldsymbol{A}}_\perp(\boldsymbol{k})$, the term $\boldsymbol{k}\times\boldsymbol{\nabla}_k\tilde{A}_{\perp a}(\boldsymbol{k})$ appearing in Ref.[1], Complement $\mathrm{B_I}$, Eq.(11) can be simplified, as follows:

$$\begin{aligned}
\boldsymbol{k}\times\boldsymbol{\nabla}_k\tilde{A}_{\perp a}(\boldsymbol{k}) &= \boldsymbol{k}\times\boldsymbol{\nabla}_k(-\mathrm{i}\omega^{-1}\tilde{E}_a) = -\mathrm{i}c^{-1}\boldsymbol{k}\times\boldsymbol{\nabla}_k(k^{-1}\tilde{E}_a) = -\mathrm{i}c^{-1}\boldsymbol{k}\times[(\boldsymbol{\nabla}_k k^{-1})\tilde{E}_a + k^{-1}\boldsymbol{\nabla}_k\tilde{E}_a]\\
&= -\mathrm{i}c^{-1}\boldsymbol{k}\times[-(\boldsymbol{k}/k^3)\tilde{E}_a + k^{-1}\boldsymbol{\nabla}_k\tilde{E}_a] = -\mathrm{i}\omega^{-1}\boldsymbol{k}\times\boldsymbol{\nabla}_k\tilde{E}_a(\boldsymbol{k}).
\end{aligned}$$

$$\begin{aligned}\boldsymbol{\mathcal{L}} &= -\mathrm{i}\varepsilon_0 \iiint_{-\infty}^{\infty} \omega^{-1}\{(ck_z/k)^2[(\boldsymbol{\mathcal{E}}\cdot\partial_{k_x}\boldsymbol{\mathcal{E}}^*)\hat{\boldsymbol{x}} + (\boldsymbol{\mathcal{E}}\cdot\partial_{k_y}\boldsymbol{\mathcal{E}}^*)\hat{\boldsymbol{y}}] - (c/k)^2(\boldsymbol{\mathcal{E}}\cdot\boldsymbol{\mathcal{E}}^*)(k_x\hat{\boldsymbol{x}} + k_y\hat{\boldsymbol{y}})\} \times \boldsymbol{k} \\ &\qquad \times (ck_z/k)^{-1}\mathrm{d}k_x\mathrm{d}k_y\mathrm{d}\omega \\ &= -\mathrm{i}\varepsilon_0 \iiint_{-\infty}^{\infty} (c/\omega)^2\big[k_z(\boldsymbol{\mathcal{E}}\cdot\partial_{k_x}\boldsymbol{\mathcal{E}}^*)\hat{\boldsymbol{x}} + k_z(\boldsymbol{\mathcal{E}}\cdot\partial_{k_y}\boldsymbol{\mathcal{E}}^*)\hat{\boldsymbol{y}} + (\boldsymbol{\mathcal{E}}\cdot\boldsymbol{\mathcal{E}}^*)\hat{\boldsymbol{z}}\big] \times \boldsymbol{k}\,\mathrm{d}k_x\mathrm{d}k_y\mathrm{d}\omega \\ &= -\mathrm{i}\varepsilon_0 \iiint_{-\infty}^{\infty} (c/\omega)^2 k_z\big[(\boldsymbol{\mathcal{E}}\cdot\partial_{k_x}\boldsymbol{\mathcal{E}}^*)\hat{\boldsymbol{x}} + (\boldsymbol{\mathcal{E}}\cdot\partial_{k_y}\boldsymbol{\mathcal{E}}^*)\hat{\boldsymbol{y}}\big] \times \boldsymbol{k}\,\mathrm{d}k_x\mathrm{d}k_y\mathrm{d}\omega. \qquad \text{(A12)}\end{aligned}$$

Real-valued term, having odd symmetry with respect to the origin, integrates to zero

Once again, the last expression is seen to agree with the one appearing within Eq.(12) of our *Phys. Rev. A* (2011) paper,[6] except for the leading minus sign, which was missed in our original paper. This minus sign was overlooked when $\partial_{k_x}\boldsymbol{\mathcal{E}}(-k_x, -k_y, -\omega)$ and $\partial_{k_y}\boldsymbol{\mathcal{E}}(-k_x, -k_y, -\omega)$ were replaced by $\partial_{k_x}\boldsymbol{\mathcal{E}}^*(k_x, k_y, \omega)$ and $\partial_{k_y}\boldsymbol{\mathcal{E}}^*(k_x, k_y, \omega)$ in the final step of the calculation.